\documentclass[11pt]{article}
\usepackage{amsmath,epsfig,amsthm,amsfonts,amssymb,bm}
\usepackage[sort]{natbib}
\usepackage{siunitx}
\usepackage{blindtext}
\usepackage[T1]{fontenc}
\usepackage[utf8]{inputenc}
\usepackage{float}
\usepackage{graphicx}
\usepackage{verbatim}
\usepackage{appendix}
\usepackage{xcolor}
\usepackage{enumitem}
\usepackage{caption}
\usepackage{subcaption}
\usepackage{booktabs}
\makeatletter \@addtoreset{equation}{section}
\renewcommand{\thesection}{\arabic{section}}
\renewcommand{\theequation}{\thesection.\arabic{equation}}

\newcommand{\beq}[1]{\begin{equation}\label{#1}}
\newcommand{\eeq}{\end{equation}}
\newcommand{\bed}{\begin{displaymath}}
\newcommand{\eed}{\end{displaymath}}
\newcommand{\ben}{\begin{eqnarray*}}
\newcommand{\een}{\end{eqnarray*}}

\renewcommand{\hat}{\widehat}

\newtheorem{thm}{Theorem}[section]

\theoremstyle{definition}
\newtheorem{rem}[thm]{Remark}

\newcommand{\bedd}{\bed\begin{array}{l}}
\newcommand{\eedd}{\end{array}\eed}

\def\({\left(}
\def\){\right)}

\newcommand{\nd}{\noindent}

\def\one{{\hbox{1{\kern -0.35em}1}}}

\newcommand{\bea}{\bed\begin{array}{rl}}
\newcommand{\eea}{\end{array}\eed}
\newcommand{\barray}{\begin{array}{ll}}
\newcommand{\earray}{\end{array}}

\begin{document}

\date{}
\title{A comparative analysis of several multivariate zero-inflated and zero-modified models with applications in insurance}
\author{Pengcheng Zhang\thanks{School of Insurance, Shandong University of Finance and Economics, Jinan 250014, China}
\and David Pitt\thanks{Centre for Actuarial Studies, Department of Economics, The University of Melbourne, VIC 3010, Australia}
\and Xueyuan Wu\thanks{Corresponding author. Email: xueyuanw@unimelb.edu.au. Centre for Actuarial Studies, Department of Economics, The University of Melbourne, VIC 3010, Australia}
}
\maketitle

\begin{abstract}

Claim frequency data in insurance records the number of claims on insurance policies during a finite period of time. Given that insurance companies operate with multiple lines of insurance business where the claim frequencies on different lines of business are often correlated, multivariate count modeling with dependence for claim frequency is therefore essential. Due in part to the operation of bonus-malus systems, claims data in automobile insurance are often characterized by an excess of common zeros. This feature is referred to as multivariate zero-inflation. In this paper, we establish two ways of dealing with this feature. The first is to use a multivariate zero-inflated model, where we artificially augment the probability of common zeros based on standard multivariate count distributions. The other is to apply a multivariate zero-modified model, which deals with the common zeros and the number of claims incurred in each line given that at least one claim occurs separately. A comprehensive comparative analysis of several models under these two frameworks is conducted using the data of an automobile insurance portfolio from a major insurance company in Spain. A less common situation in insurance is the absence of some common zeros resulting from incomplete records. This feature of these data is known as multivariate zero-deflation. In this case, our proposed multivariate zero-modified model still works, as shown by the second empirical study.

\vskip 0.25 in \nd{\textbf{Key Words}: Multivariate zero-inflated model; Multivariate zero-modified model; EM algorithm; MM algorithm; Automobile insurance; Ratemaking}
\end{abstract}

\section{Introduction}

Ratemaking is one of the main tasks that actuaries perform in the automobile insurance sector. Designing a proper tariff structure for an insurance company is a key actuarial task. Therefore, many attempts have been made in the actuarial literature to find an appropriate distribution for the annual number of claims filed by insurance policyholders. A thorough review of ratemaking procedures when modeling claim count data in automobile insurance can be found in \cite{Denuit-Marechal-Pitrebois-Walhin-2007}. Claim frequency data in automobile insurance is often characterized by a large number of zero claims. This is because the insureds often do not experience an accident or other claim-related event during their period of insurance coverage and also may be reluctant to report small claims so as to maintain a high level of no-claim discount (NCD) on their future insurance premiums.

There are two commonly used methods to cope with the excess of zeros in claim counts. The most popular one is using a zero-inflated model, which can be considered as a mixture of a point mass at zero and a standard count distribution. The application of several zero-inflated models in automobile insurance can be found in \cite{Yip-Yau-2005}. Another approach is to apply a hurdle model (\cite{Mullahy-1986}), which uses a different mechanism to modify a basic count distribution to represent the situation with excess zeros. It is motivated by a two-stage decision-making process confronted by individuals. The superiority of the hurdle model is that it can flexibly deal with both zero-deflation and zero-inflation phenomena. A comparative analysis of some zero-inflated models and hurdle models in automobile insurance can be found in \cite{Boucher-Denuit-Guillen-2007}.

In practice, it is not uncommon to model claim counts for different types of claims on an insurance policy separately. Actuaries typically assume independence between the number of each type of claim when faced with the challenge of pricing an insurance contract with multiple types of coverage. In many circumstances, the accuracy of this assumption is questionable. Several studies have revealed the risks associated with ignoring the dependence between different types of claims, see, e.g., \cite{Frees-Valdez-2008} and \cite{Frees-Shi-Valdez-2009}.

The multivariate Poisson model is a frequently employed model in insurance that takes correlations between multivariate counts into account. This model is constructed from the summation of independent Poisson random variables. The dependence is introduced with shared variables (also known as a common shock). The application of this multivariate Poisson model in an actuarial setting can be found in \cite{Bermudez-2009} and \cite{Bermudez-Karlis-2011}. Another commonly used model is the multivariate mixed Poisson model (see \cite{Ghitany-Karlis-Mutairi-Awadhi-2012}), which is based on a mixture of independent Poisson random variables. A critical property of this mixed Poisson model is its ability to account for overdispersion. 

Similar to the univariate case, the claims data in automobile insurance often exhibit zero-inflation in the multivariate context. In the literature, most papers have dealt with this feature under the framework of zero-inflated models. The main idea behind these models is to mix a multivariate count distribution with a point mass at zero for all components of the multivariate random variable. \cite{Bermudez-Karlis-2011} considered a multivariate zero-inflated Poisson model where the multivariate Poisson distribution is constructed using common shock variables. \cite{Liu-Tian-2015} examined a multivariate zero-inflated Poisson model derived from independent Poisson distributions. \cite{Zhang-Pitt-Wu-2022} further proposed a multivariate zero-inflated hurdle model where each margin independently follows a zero-modified distribution instead of a simple Poisson variable. This significantly enhances the model's flexibility.

A less developed approach to addressing the issue of multivariate zero-inflation is to generalize the zero-modified model from the univariate case to the multivariate case. The key idea is to separate the common zeros from the data and deal with them individually. The resultant model then can flexibly deal with both multivariate zero-inflation and zero-deflation phenomena. \cite{Tian-Liu-Tang-Jiang-2018} developed a multivariate zero-modified Poisson model starting from independent Poisson random variables. \cite{Liu-Tian-Tang-Yuen-2019} investigated a multivariate zero-modified Poisson model based on the common shock multivariate Poisson.

As for the inference, we apply the expectation-maximization (EM) algorithm (\cite{Dempster-Laird-Rubin-1977}) for those multivariate zero-inflated models. The EM algorithm is a two-step iterative method to find the maximum likelihood estimates (MLEs). It is particularly useful when working with zero-inflated models. Examples illustrating the implementation of the EM algorithm in zero-inflated models can be found in \cite{Lambert-1992}, \cite{Hall-2000} in the univariate case, and \cite{Liu-Tian-2015}, \cite{Zhang-Pitt-Wu-2022} in the multivariate case. 

For multivariate zero-modified models, the observed log-likelihood function can be split into two parts that can be optimized separately. The estimation regarding the zero-modification parameter can be accomplished via logistic regression. Then the problem reduces to the estimation of the parameters in a multivariate zero-truncated model. Previous papers addressed this issue with the help of an EM-type algorithm (see e.g., \cite{Tian-Liu-Tang-Jiang-2018}, \cite{Liu-Tian-Tang-Yuen-2019}, \cite{Zhang-Calderin-Li-Wu-2020}). However, we find that this EM-type algorithm fails to converge to the MLEs when covariates are introduced. Thus, we adopt a minorization-maximization (MM) algorithm for estimation. The first M-step is to construct a surrogate function that minorizes the objective function. The second M-step is to maximize the surrogate function. Usually, the surrogate function has a simpler form and is easier to maximize compared with the objective function. We iteratively carry out the two steps until convergence. A thorough introduction to the MM algorithm can be found in \cite{Hunter-Lange-2004}. The usefulness of the MM algorithm in univariate zero-truncated models can be found in \cite{Zhou-Lange-2010}.

Every EM algorithm can be regarded as an example of the MM algorithm. In the E-step of the EM algorithm, we need to calculate the $Q$ function, which is the conditional expectation of the complete log-likelihood. It can be shown that this $Q$ function in the E-step is, up to a constant, a minorizing function of the objective function. Thus, we can take advantage of this point to construct the surrogate function in the first M-step in the MM algorithm.

Our work contributes to the existing literature in several ways. First, we offer two effective strategies for handling multivariate zero-inflated data in insurance. Several models are proposed under these two frameworks. To our knowledge, some of the models explored have never been considered in an insurance-related setting. Second, we develop a complete set of inference tools for these models, including EM and MM algorithms. Compared with the EM algorithm, the MM algorithm has seldom been applied in the actuarial literature. Third, we compare these models from a number of perspectives with the help of an automobile insurance data set. This provides some insight for practitioners in the insurance industry.

The rest of this paper is organized as follows. Section 2 reviews several commonly used multivariate count models in insurance. Sections 3 and 4 propose several multivariate zero-inflated and zero-modified models derived from those put forward in Section 2 via stochastic representation. The detailed inference procedures are also given. In Section 5, we compare these proposed models based on a real insurance data set. Section 6 concludes the paper.

\section{Review of several multivariate count models}
In this section, some commonly used multivariate count models in insurance are reviewed. We denote $\bm{Y}=(Y_1, \ldots$, $Y_m)^\top$ as a discrete random vector where $Y_j$, $j=1,\ldots, m$, denotes the number of claims of type $j$.

\subsection{Multivariate independent count models}
The most common method to model multivariate counts is to neglect the correlations between the type of claims and build models independently based on each type. Possible choices for margins include Poisson, negative binomial and their zero-inflated and hurdle versions.

\subsubsection{Poisson model}
The probability mass function (pmf) of the Poisson distribution is given by
\begin{eqnarray}
\Pr(Y=y)=\frac{\lambda^{y}}{y!}e^{-\lambda}.
\end{eqnarray}
One underlying assumption of the Poisson distribution is equidispersion, meaning that $\mbox{E}(Y)=\mbox{Var}(Y)$.

\subsubsection{Negative binomial model}
It is often the case that the unobserved heterogeneity in
the insurance data will lead to overdispersion. This cannot
be fully remedied by a simple Poisson model. To account for overdispersion, we can use the negative distribution. The pmf of the negative binomial distribution is given by
\begin{eqnarray}
\Pr(Y=y)=\frac{\Gamma(y+\phi)}{\Gamma(\phi)y!} \left(\frac{\lambda}{\lambda+\phi}\right)^{y}
\left(\frac{\phi}{\lambda+\phi}\right)^{\phi}.
\end{eqnarray}

\subsubsection{Zero-inflated model}
The zero-inflated count model provides a way to model count data with excess zeros. Let $Y$ follow a standard count distribution defined on $\mathbb{N}$, then $Z$ is said to follow the zero-inflated distribution if
\begin{eqnarray}
Z \overset{d}{=} U_0 Y = \left\{ \begin{gathered}
  0, \quad \hfill U_0=0,  \\
  Y, \quad \hfill U_0=1,  \\
\end{gathered}  \right.
\end{eqnarray}
where $U_0 \sim {\rm Bernoulli}(\pi_0)$, $0<\pi_0<1$, and $U_0$ is independent of $Y$. The symbol $``\overset{d}{=}"$ means that the random variables on both sides of the equality share the same distribution. The pmf of $Z$ can be derived as
\begin{eqnarray}
\Pr(Z=z)= \left\{ \begin{gathered}
  1-\pi_0+\pi_0\Pr(Y=0), \quad \hfill z=0,  \\
  \pi_0 \Pr(Y=z), \quad \hfill z>0.  \\
\end{gathered}  \right.
\end{eqnarray}
Commonly used distributions for $Y$ include Poisson and negative binomial.

\subsubsection{Zero-modified (Hurdle) model}
The zero-modified count model is a two-part model that separates the occurrence of an event from the number of those events actually observed. To construct a zero-modified distribution, we first provide the definition for a zero-truncated distribution. Let $Y$ follow a standard count distribution defined on $\mathbb{N}$, then $W$ is said to follow a zero-truncated distribution if
\begin{eqnarray}
Y \overset{d}{=} U_0 W=\left\{ \begin{gathered}
  0, \quad \hfill U_0=0,  \\
  W, \quad \hfill U_0=1,  \\
\end{gathered}  \right. 
\end{eqnarray}
where $U_0 \sim {\rm Bernoulli}(\pi_0)$, $\pi_0=\Pr(Y \neq 0)$, and $U_0$ is independent of $W$. The pmf of $W$ can be derived as
\begin{eqnarray}
\Pr(W=w)=\frac{\Pr(Y=w)}{\pi_0},\quad w>0.
\end{eqnarray}

We now define the zero-modified distribution. $Z$ is said to follow a zero-modified distribution if
\begin{eqnarray}
Z \overset{d}{=} U'_0 W = \left\{ \begin{gathered}
  0, \quad \hfill U'_0 = 0,  \\
  W, \quad \hfill U'_0 = 1,  \\
\end{gathered}  \right. 
\end{eqnarray}
where $U'_0 \sim {\rm Bernoulli}(\pi'_0)$, $0<\pi'_0 < 1$, and $U'_0$ is independent of $W$. The pmf of $\bm{Z}$ can be derived as
\begin{eqnarray}
\Pr(Z=z)=\left\{ \begin{gathered}
  1-\pi'_0 , \quad \hfill z=0,  \\
  \frac{\pi'_0}{\pi_0}  \Pr(Y=z), \quad \hfill z>0.  \\
\end{gathered}  \right.
\label{UZM}
\end{eqnarray}

Several special cases of (\ref{UZM}) are given as follows.
\begin{itemize}
\item $0<\pi'_0<\pi_0$, $Z$ follows a zero-inflated count distribution.
\item $\pi'_0=\pi_0$, $Z \overset{d}{=} Y$, $Z$ follows a standard count distribution..
\item $\pi_0<\pi'_0<1$, $Z$ follows a zero-deflated count distribution.
\end{itemize}

\begin{rem}
An alternative choice for $W$ is the unit-shifted distribution. The pmf of $W$ can be defined as follows: $\Pr(W=w)=\Pr(Y=w-1)$, $w>0$.
\end{rem}

Throughout the paper, we consider four potential choices for $W$: zero-truncated Poisson (ZTP), zero-truncated negative binomial (ZTNB), unit-shifted Poisson (USP) and unit-shifted negative binomial (USNB).

\subsection{Multivariate Poisson model}
We take advantage of common shocks to construct the multivariate Poisson distribution. Let
\begin{eqnarray}
Y_j=N_j+N_0, \quad j=1,\ldots,m,
\end{eqnarray}
where each $N_j$, $j=0,\ldots,m$, independently follows a simple Poisson distribution with parameter $\lambda_j$. Then $\bm{Y}=(Y_1, \ldots$, $Y_m)^\top$ is said to follow the multivariate Poisson distribution. The pmf of $\bm{Y}$ is given by
\begin{eqnarray}
\Pr(\bm{Y}=\bm{y})=e^{-\lambda'}\sum_{n_0=0}^{l}\left[\frac{\lambda_0^{n_0}}{n_0!}
\prod_{j=1}^{m}\frac{\lambda_j^{y_j-n_0}}{(y_j-n_0)!}\right],
\label{MP}
\end{eqnarray}
where $\lambda'=\sum_{j=0}^{m}\lambda_j$ and $l=\min(y_1,\ldots,y_m)$. When $\lambda_0=0$, the multivariate Poisson distribution simplifies to the independent case. 

\subsection{Multivariate negative binomial model}
Let each $Y_j$ independently follow a Poisson distribution with parameter $\alpha \lambda_j$ where $\alpha$ follows a gamma distribution with shape and rate parameters both equal to $\phi$. In this case, $\bm{Y}$ is said to follow the multivariate negative binomial distribution. The pmf of $\bm{Y}$ is given by
\begin{eqnarray}
\Pr(\bm{Y}=\bm{y}) \hspace{-0.1in} &=& \hspace{-0.1in} 
\int_0^\infty \prod_{j=1}^{m} \frac{\exp ( -\alpha \lambda_j) (\alpha \lambda _j)^{y_j}} {{y_j}!}g(\alpha; \phi) d\alpha, 
\end{eqnarray}
where
\begin{eqnarray}
g(\alpha;\phi)=\frac{\phi^{\phi}}{\Gamma(\phi)} \alpha^{\phi-1}e^{-\phi\alpha}.
\end{eqnarray}
The pmf of $\bm{Y}$ can be written in the closed form:
\begin{eqnarray}
\Pr(\bm{Y}=\bm{y}) \hspace{-0.1in} &=& \hspace{-0.1in} \frac{\Gamma(\sum_{j=1}^{m}y_j+\phi)}{\Gamma(\phi) \prod_{j=1}^{m}y_j!}\frac{\phi^\phi \prod_{j=1}^{m}\lambda_j^{y_j}}{(\phi+\sum_{j=1}^{m} \lambda_j)^{\sum_{j=1}^{m}y_j+\phi}}.
\label{MNB}
\end{eqnarray}


\section{Multivariate zero-inflated models}

\subsection{Definition}
Let $ \bm{Y}=(Y_1, \ldots,Y_m)^\top $ denote a discrete random vector where each $Y_j$ is defined on $\mathbb{N}$. Then $\bm{Z}=(Z_1,\ldots,Z_m)^\top$ is said to follow a multivariate zero-inflated distribution if
\begin{eqnarray}
\bm{Z} \overset{d}{=} U_0\bm{Y} = \left\{ \begin{gathered}
  \bm{0}, \quad \hfill U_0 = 0,  \\
  \bm{Y}, \quad \hfill U_0 = 1,  \\
\end{gathered}  \right.\label{MZIH-M1}
\end{eqnarray}
where $U_0 \sim {\rm Bernoulli}(\pi_0)$, $0<\pi_0<1$, and $U_0$ is independent of $\bm{Y}$. The pmf of $\bm{Z}$ can be derived as
\begin{eqnarray}
\Pr(\bm{Z}=\bm{z})=
  \left[1-\pi_0+\pi_0\Pr(\bm{Y}=0)\right]^{v}\left[\pi_0 \Pr(\bm{Y}=\bm{z})\right]^{1-v},
\end{eqnarray}
where $\bm{z}=(z_1,\ldots,z_m)^\top$ is a vector of observed values, $v=\mathbb{I}(\bm{z}=\bm{0})$ and $\mathbb{I}(\cdot)$ is an indicator function.

The covariates for $\pi_0$ can be incorporated via a logit-link function:
\begin{eqnarray}
\pi_0=\frac{\exp(\bm{x}^\top \bm{\gamma})}{1+\exp(\bm{x}^\top \bm{\gamma})}, 
\end{eqnarray}
where $\bm{x}=(1, x_{1},\ldots, x_{p})^\top$ and $\bm{\gamma}=(\gamma_{0},\gamma_{1},\ldots,\gamma_{p})^\top$.

\subsection{The models}
\subsubsection{Type I multivariate zero-inflated Poisson model}
Let each $Y_j$ independently follow a Poisson distribution with parameter $\lambda_j$. Then $\bm{Z}$ is said to follow the Type I multivariate zero-inflated Poisson distribution (MZIP) with parameters $\bm{\lambda}=(\lambda_1,\ldots, \lambda_m)^\top$ and  $\pi_0$, denoted by $\bm{Z} \sim {\rm MZIP^{(I)}}(\bm{\lambda},\pi_0)$. The pmf of $\bm{Z}$ is
\begin{eqnarray*}
\Pr(\bm{Z}=\bm{z})=
  \left(1-\pi_0+\pi_0e^{-\sum_{j=1}^{m}\lambda_j}\right)^{v}
\left(\pi_0\prod_{j=1}^{m}
  \frac{\lambda_j^{z_j}e^{-\lambda_j}}{z_j!}\right)^{1-v}.
\end{eqnarray*}
The covariates can be incorporated via a log-link function:
\begin{eqnarray}
\lambda_{j}=\exp({\bm{x}^\top \bm{\beta}_j}), \quad j=1\ldots,m,
\end{eqnarray}
where $\bm{x}=(1, x_{1},\ldots, x_{p})^\top$ and $\bm{\beta}_j=(\beta_{j0},\beta_{j1},\ldots,\beta_{jp})^\top$.

\subsubsection{Type I multivariate zero-inflated negative binomial model}
Let each $Y_j$ independently follow a negative binomial distribution with parameters $\lambda_j$ and $\phi_j$. Then $\bm{Z}$ is said to follow the Type I multivariate zero-inflated negative binomial (MZINB) distribution with parameters $\bm{\lambda}=(\lambda_1,\ldots, \lambda_m)^\top$, $\bm{\phi}=(\phi_1,\ldots, \phi_m)^\top$ and $\pi_0$, denoted by $\bm{Z} \sim {\rm MZINB^{(I)}}(\bm{\lambda},\bm{\phi},\pi_0)$. The pmf of $\bm{Z}$ is
\begin{eqnarray*}
\Pr(\bm{Z}=\bm{z}) \hspace{-0.1in} &=& \hspace{-0.1in} \left[1-\pi_0+\pi_0\prod_{j=1}^m{\left(\frac{\phi_j}{\lambda_j+\phi_j} \right)}^{\phi_j}\right]^{v} \nonumber\\
 && \hspace{-0.1in} \times \left[\pi_0\prod_{j=1}^{m} \frac{\Gamma({z_j}+{\phi_j})}{\Gamma({\phi_j}){z_j}!} \left(\frac{\lambda_j}{\lambda_j+\phi_j}\right)^{z_j}
 \left(\frac{\phi_j}{\lambda_j+\phi_j}\right)^{\phi_j}\right]^{1-v}.
\end{eqnarray*}
The covariates can be incorporated via a log-link function:
\begin{eqnarray}
\lambda_{j}=\exp({\bm{x}^\top \bm{\beta}_j}), \quad j=1\ldots,m,
\end{eqnarray}
where $\bm{x}=(1, x_{1},\ldots, x_{p})^\top$ and where $\bm{\beta}_j=(\beta_{j0},\beta_{j1},\ldots,\beta_{jp})^\top$.

\subsubsection{Type I multivariate zero-inflated hurdle model}
Let each $Y_j$ independently follow a zero-modified distribution, which can be characterized as follows:
\begin{eqnarray}
Y_j \overset{d}{=} U_j W_j = \left\{ \begin{gathered}
  0, \quad \hfill U_j = 0,  \\
  W_j, \quad \hfill U_j = 1,  \\
\end{gathered}  \right. 
\end{eqnarray}
where $W_j$ follows a count distribution defined on $\mathbb{N}_{+}$, $U_j \sim {\rm Bernoulli}(\pi_j)$, $0<\pi_j<1$, and $U_j$ is independent of $W_j$. Then $\bm{Z}$ is said to follow the Type I multivariate zero-inflated hurdle (MZIH) distribution with parameters $\bm{\pi}=(\pi_1,\ldots, \pi_m)^\top$, $\bm{\Omega}=(\bm{\Omega}_1,\ldots, \bm{\Omega}_m)^\top$ and $\pi_0$, denoted by $\bm{Z} \sim {\rm MZIH^{(I)}}(\bm{\pi},\bm{\Omega},\pi_0)$. Here $\bm{\Omega}_j$ is the set of parameters related to $W_j$. The pmf of $\bm{Z}$ is
\begin{eqnarray}
\Pr(\bm{Z}=\bm{z}) \hspace{-0.1in}&=&\hspace{-0.1in}
  \left[1-\pi_0+\pi_0\prod_{j=1}^{m}(1-\pi_j)\right]^{v} \nonumber \\
 &&\hspace{-0.1in} \times
  \left[\pi_0 \prod_{j:z_j=0}(1-\pi_j) \prod_{j:z_j \ne 0}\pi_j f_{W_j}(z_j) \right]^{1-v}.
\end{eqnarray}
The covariates for $\pi_j$ can be incorporated via a logit-link function:
\begin{eqnarray}
\pi_j=\frac{\exp(\bm{x}^\top \bm{\beta}_j)}{1+\exp(\bm{x}^\top \bm{\beta}_j)}, \quad j=1\ldots,m,
\end{eqnarray}
where $\bm{x}=(1, x_{1},\ldots, x_{p})^\top$ and where $\bm{\beta}_j=(\beta_{j0},\beta_{j1},\ldots,\beta_{jp})^\top$. The covariates for the location parameter $\lambda_j$ of $W_j$ can be incorporated via a log-link function:
\begin{eqnarray}
\lambda_{j}=\exp({\bm{x}^\top \bm{\alpha}_j}), \quad j=1\ldots,m,
\end{eqnarray}
where $\bm{x}=(1, x_{1},\ldots, x_{p})^\top$ and  $\bm{\alpha}_j=(\alpha_{j0},\alpha_{j1},\ldots,\alpha_{jp})^\top$. 


\subsubsection{Type II multivariate zero-inflated Poisson  model}
Let $\bm{Y}$ follow the multivariate Poisson distribution given in (\ref{MP}). Then $\bm{Z}$ is said to follow the Type II multivariate zero-inflated Poisson (MZIP) distribution with parameters $\lambda_0$, $\bm{\lambda}=(\lambda_1,\ldots, \lambda_m)^\top$ and $\pi_0$, denoted by $\bm{Z} \sim {\rm MZIP^{(II)}}(\lambda_0,\bm{\lambda},\pi_0)$. The pmf of $\bm{Z}$ is
\begin{eqnarray}
\Pr(\bm{Z}=\bm{z})=\left(1-\pi_0+\pi_0 e^{-\lambda'}\right)^{v} 
\left\{\pi_0e^{-\lambda'}\sum_{n_0=0}^{l}\left[\frac{\lambda_0^{n_0}}{n_0!}
\prod_{j=1}^{m}\frac{\lambda_j^{z_j-n_0}}{(z_j-n_0)!}\right]\right\}^{1-v}.
\end{eqnarray}
When $\lambda_0=0$, the Type II MZIP distribution simplifies to the Type I case. The covariates can be incorporated via a log-link function:
\begin{eqnarray}
\lambda_{j}=\exp({\bm{x}^\top \bm{\beta}_j}), \quad j=1\ldots,m,
\end{eqnarray}
where $\bm{x}=(1, x_{1},\ldots, x_{p})^\top$ and  $\bm{\beta}_j=(\beta_{j0},\beta_{j1},\ldots,\beta_{jp})^\top$. For ease of interpretation, we do not incorporate covariates in $\lambda_0$.

\subsubsection{Type II multivariate zero-inflated negative binomial model}

Let $\bm{Y}$ follow the multivariate negative binomial distribution given in (\ref{MNB}). Then $\bm{Z}$ is said to follow the Type II multivariate zero-inflated negative binomial (MZINB) distribution with parameters $\bm{\lambda}=(\lambda_1,\ldots, \lambda_m)^\top$, $\phi$ and $\pi_0$, denoted by $\bm{Z} \sim {\rm MZINB^{(II)}}(\bm{\lambda},\phi,\pi_0)$. The pmf of $\bm{Z}$ is
\begin{eqnarray}
\Pr(\bm{Z}=\bm{z}) \hspace{-0.1in} &=& \hspace{-0.1in} \left[1-\pi_0+\pi_0\left(\frac{\phi}{\sum_{j=1}^{m} \lambda_j+\phi} \right)^{\phi}\right]^{v} \nonumber\\
 && \hspace{-0.1in} \times 
\left[\pi_0  \frac{\Gamma\left(\sum_{j=1}^{m}z_j+\phi\right)}{\Gamma(\phi) \prod_{j=1}^{m}z_j!}\frac{\phi^\phi \prod_{j=1}^{m}\lambda_j^{z_j}}{\left(\sum_{j=1}^{m} \lambda_j+\phi \right)^{\sum_{j=1}^{m}z_j+\phi}} \right]^{1-v}. 
\end{eqnarray}
The covariates can be incorporated via a log-link function:
\begin{eqnarray}
\lambda_{j}=\exp({\bm{x}^\top \bm{\beta}_j}), \quad j=1\ldots,m,
\end{eqnarray}
where $\bm{x}=(1, x_{1},\ldots, x_{p})^\top$ and  $\bm{\beta}_j=(\beta_{j0},\beta_{j1},\ldots,\beta_{jp})^\top$.

\subsection{Inference}
Suppose we have a sample of size $n$. The corresponding observed values are $\bm{z}_1,\ldots, \bm{z}_n$, where $\bm{z}_i=(z_{i1},\ldots, z_{im})^\top$. The indicator variables are $v_1 ,\ldots, v_n$, where $v_i=\mathbb{I}(\bm{z}_i=\bm{0})$. Covariates are $\bm{x}_1,\ldots, \bm{x}_n$, where $\bm{x}_i=(1, x_{i1},\ldots, x_{ip})^\top$. The likelihood function then can be written as
\begin{eqnarray}
L(\bm{\Theta}) \hspace{-0.1in}&=&\hspace{-0.1in} \prod_{i=1}^n \left[1-\pi_{0i}+\pi_{0i} f_{\bm{Y}}(\bm{0})\right]^{v_i} \prod_{i=1}^n\left[\pi_{0i} f_{\bm{Y}}(\bm{z_i})\right]^{1-v_i},
\end{eqnarray}
where $\pi_{0i}=\frac{\exp(\bm{x}_i^\top \bm{\gamma})}{1+\exp(\bm{x}_i^\top \bm{\gamma})}$ and $v_i=\mathbb{I}(\bm{z}_i=\bm{0})$. $\bm{\Theta}$ is the total set of parameters to estimate.

\subsubsection{Case 1-3: Type I MZIP, MZINB, MZIH models}
For the three Type I cases, the log-likelihood function $\ell$ can be written as:
\begin{eqnarray*}
\ell(\bm{\Theta}) 
\hspace{-0.1in}&=&\hspace{-0.1in}
\sum_{i=1}^n v_i \log \left[1-\pi_{0i}+\pi_{0i} \prod_{j=1}^m f_{Y_j}(0)\right]+ \sum_{i=1}^n (1-v_i) \log \pi_{0i}\\
&&\hspace{-0.1in} 
+\sum_{i=1}^n \sum_{j=1}^m (1-v_i) f_{Y_j}(z_{ij}).
\end{eqnarray*}
We can implement the EM algorithm as proposed in \cite{Zhang-Pitt-Wu-2022} for these three cases.

\subsubsection{Case 4: Type II MZIP model}
Suppose we observe the values $u'_i$ and $n_{0i}$, where $u'_i=1$ indicates the observation of common zeros is inflated and 0 otherwise, then the complete log-likelihood function of $\ell^c$ can be written as follows.
\begin{eqnarray*}
{\ell^c}(\bm{\Theta}) 
\hspace{-0.1in} &\propto& \hspace{-0.1in} 
\sum_{i=1}^n \left[u'_i v_i\log(1-\pi_{0i})+(1-u'_i v_i)\log\pi_{0i}\right] \\ 
&&\hspace{-0.1in}
+\sum_{i=1}^n \left[n_{0i}\log \lambda_0 - (1-u'_i v_i)\lambda_0 \right] \\
&&\hspace{-0.1in}
+\sum_{i=1}^n \sum_{j=1}^{m}   \left[ (z_{ij}-n_{0i}) \log \lambda_{ij}-(1-u'_i v_i)\lambda_{ij} \right],
\end{eqnarray*}
where $\lambda_{ij}=\exp({\bm{x}_{i}^\top \bm{\beta}_j})$. 

The $Q$ function at the $t$-th iteration is given by
\begin{eqnarray*}
Q(\bm{\Theta}; \bm{\Theta}^{(t)})
\hspace{-0.1in} &=& \hspace{-0.1in} 
\sum_{i=1}^n \left[ u_i^{\prime(t)} v_i \log(1-\pi_{0i}) + \left( 1-u_i^{\prime(t)} v_i \right) \log \pi_{0i}\right] \\
&&\hspace{-0.1in}
+\sum_{i=1}^n \left[ n_{0i}^{(t)} \log \lambda_0 - \left( 1-u_i^{\prime(t)} v_i \right) \lambda_0 \right] \\
&&\hspace{-0.1in}
+\sum_{i=1}^n \sum_{j=1}^{m}   \left[ \left( z_{ij} - n_{0i}^{(t)} \right) \log\lambda_{ij} - \left( 1-u_i^{\prime(t)} v_i \right) \lambda_{ij} \right].
\end{eqnarray*}

\begin{itemize}
\item
E-step: 
\begin{itemize}
\item
The conditional expectation $u_i^{\prime(t)}$ is given by
\begin{eqnarray*}
u_i^{\prime(t)} \hspace{-0.1in}&=&\hspace{-0.1in} \mbox{E}_{u'}\left(u'_i|\bm{z}_i=\bm{0},\bm{\Theta}^{(t)}\right)=
\frac{1-\pi_{0i}^{(t)}}{1-\pi_{0i}^{(t)}+\pi_{0i}^{(t)} e^{-\lambda_0^{(t)}-\sum_{j=1}^m \lambda_{ij}^{(t)}}}.
\end{eqnarray*}
\item
The condition expectation $n_{0i}^{(t)}$ is given by 
\begin{eqnarray*}
n_{0i}^{(t)} \hspace{-0.1in}&=&\hspace{-0.1in}
\mbox{E}_{n_0}\left(n_{0i}|\bm{z}_i,\bm{\Theta}^{(t)} \right)\\
\hspace{-0.1in}&=&\hspace{-0.1in} 
\left\{ \begin{gathered}
 \frac{\lambda_0^{(t)} f_{\bm{Y}}\left(\bm{z}_i-\bm{1} | \bm{\Theta}^{(t)}\right)}
{f_{\bm{Y}}\left(\bm{z}_i | \bm{\Theta}^{(t)}\right)}, \quad \hfill \min(z_{i1}, \ldots,z_{im}) > 0,  \\
  0, \quad \hfill\min(z_{i1}, \ldots,z_{im}) = 0,  \\
\end{gathered}  \right.
\end{eqnarray*}
where $\bm{1}=(1,\ldots,1)^\top$ denotes a vector of dimension $m$ with all elements equal to 1.
\end{itemize}

\item
M-step:
\begin{itemize}
\item
Update the parameter vector $\bm{\gamma}$ by implementing Newton-Raphson method for one step. The first and second order derivatives of $Q$ with respect to $\bm{\gamma}$   are given as follows:
\begin{eqnarray*}
\frac{\partial Q}{\partial \bm{\gamma}} \hspace{-0.1in}&=&\hspace{-0.1in} 
\sum_{i=1}^{n}  \left( \tau_i^{(t)} - \pi_{0i} \right) \bm{x_{i}},\\
\frac{\partial^2 Q}{\partial \bm{\gamma}\partial \bm{\gamma}^\top}  
\hspace{-0.1in}&=&\hspace{-0.1in} 
-\sum_{i=1}^{n}  \pi_{0i} \left( 1 - \pi_{0i} \right) \bm{x_{i}}\bm{x_{i}}^\top,
\end{eqnarray*}
where $\tau_i^{(t)}=1 - u_i^{\prime(t)} v_i$.
\item
Update the parameter $\lambda_0$ using the following equation:
\begin{eqnarray*}
\lambda_0^{(t+1)}=\frac{\sum_{i=1}^n n_{0i}^{(t)}}{\sum_{i=1}^n \tau_i^{(t)}}.
\end{eqnarray*}
\item
Update the parameter vector $\bm{\beta}_j$, $j=1,\ldots,m$, separately, by implementing the Newton-Raphson method for one step. The first and second order derivatives of $Q$ with respect to $\bm{\beta}_j$ are given as follows:
\begin{eqnarray*}
\frac{\partial Q}{\partial \bm{\beta}_j} \hspace{-0.1in}&=&\hspace{-0.1in} 
\sum_{i=1}^{n}  \left[ z_{ij} - n_{0i}^{(t)} - \tau_i^{(t)} \lambda_{ij} \right] \bm{x_{i}},\\
\frac{\partial^2 Q}{\partial \bm{\beta}_j \partial \bm{\beta}_j^\top}  
\hspace{-0.1in}&=&\hspace{-0.1in} 
-\sum_{i=1}^{n}  \tau_i^{(t)} \lambda_{ij} \bm{x_{i}}\bm{x_{i}}^\top.
\end{eqnarray*}
\end{itemize}

\end{itemize}

\subsubsection{Case 5: Type II MZINB model}
Suppose we observe the values $u'_i$ and $\alpha_i$, where $u'_i=1$ indicates the observation of common zeros is inflated and 0 otherwise, then the complete log-likelihood function of $\ell^c$ can be written as follows.
\begin{eqnarray*}
{\ell^c}(\bm{\Theta}) 
\hspace{-0.1in} &\propto& \hspace{-0.1in} 
\sum_{i=1}^{n} \left[u'_i v_i\log(1-\pi_0)+(1-u'_i v_i)\log\pi_0\right] \\
&&\hspace{-0.1in}
+ \sum_{i=1}^{n}  \sum_{j=1}^m \left[z_{ij} \log \lambda_{ij} - ( 1 - u'_i v_i) \alpha_i \lambda_{ij} \right]\\
&&\hspace{-0.1in}
+\sum_{i=1}^{n} \left( 1 - u'_i v_i \right)\left[ \phi \log \phi - \log\Gamma(\phi) + (\phi-1) \log\alpha -\phi \alpha \right],
\end{eqnarray*}
where $\lambda_{ij}=\exp({\bm{x}_{i}^\top \bm{\beta}_j})$.

The $Q$ function at the $t$-th iteration is given by
\begin{eqnarray*}
Q\left(\bm{\Theta}; \bm{\Theta}^{(t)}\right) \hspace{-0.1in}&=&\hspace{-0.1in} 
\sum_{i=1}^{n} \left[u_i^{\prime(t)} v_i \log(1-\pi_0) + \left( 1 - u_i^{\prime(t)} v_i \right)\log\pi_0\right] \\
&&\hspace{-0.1in}
+\sum_{i=1}^{n}  \sum_{j=1}^m \left[ z_{ij} \log \lambda_{ij} - \left( 1 - u_i^{\prime(t)} v_i \right) r_i^{(t)} \lambda_{ij} \right]\\
&&\hspace{-0.1in}
+\sum_{i=1}^{n} \left( 1 - u_i^{\prime(t)} v_i \right)\left[ \phi \log \phi -\log\Gamma(\phi) + (\phi-1) s_i^{(t)} -\phi r_i^{(t)}\right].
\end{eqnarray*}

\begin{itemize}
\item
E-step: 
\begin{itemize}
\item
The conditional expectation $u_i^{\prime(t)}$ is given by 
\begin{eqnarray*}
u_i^{\prime(t)} \hspace{-0.1in}&=&\hspace{-0.1in} \mbox{E}_{u'}\left(u'_i|\bm{z}_i=\bm{0},\bm{\Theta}^{(t)}\right)=
\frac{1-\pi_0^{(t)}}{1-\pi_0^{(t)}+\pi_0^{(t)} \left(\frac{\phi^{(t)}}{\sum_{j=1}^{m} \lambda_{ij}^{(t)} + \phi^{(t)}} \right)^{\phi^{(t)}}}.
\end{eqnarray*}
\item
The conditional expectation $r_i^{(t)}$ is given by 
\begin{eqnarray*}
r_i^{(t)} = \mbox{E}_\alpha \left(\alpha_i|\bm{z}_i,\bm{\Theta}^{(t)} \right)= \frac{ \sum_{j=1}^m z_{ij}+\phi^{(t)}} {\sum_{j=1}^{m} \lambda_{ij}^{(t)}+\phi^{(t)}}.
\end{eqnarray*}
\item
The conditional expectation $s_i^{(t)}$ is given by
\begin{eqnarray*}
s_i^{(t)} \hspace{-0.1in}&=&\hspace{-0.1in}
\mbox{E}_\alpha\left(\log \alpha_i|\bm{z}_i,\bm{\Theta}^{(t)} \right)= \psi\left(\sum_{j=1}^m z_{ij}+\phi^{(t)}\right)-\log\left(\sum_{j=1}^{m} \lambda_{ij}^{(t)}+\phi^{(t)}\right),
\end{eqnarray*}
where $\psi(\cdot)$ denotes the digamma function.
\end{itemize}

\item
M-step:
\begin{itemize}
\item
Update the parameter vector $\bm{\gamma}$ by implementing the Newton-Raphson method for one step. The first and second order derivatives of $Q$ with respect to $\bm{\gamma}$   are given as follows:
\begin{eqnarray*}
\frac{\partial Q}{\partial \bm{\gamma}} \hspace{-0.1in}&=&\hspace{-0.1in} 
\sum_{i=1}^{n}  \left( \tau_i^{(t)} - \pi_{0i} \right) \bm{x_{i}},\\
\frac{\partial^2 Q}{\partial \bm{\gamma}\partial \bm{\gamma}^\top}  
\hspace{-0.1in}&=&\hspace{-0.1in} 
-\sum_{i=1}^{n}  \pi_{0i} \left( 1 - \pi_{0i} \right) \bm{x_{i}}\bm{x_{i}}^\top,
\end{eqnarray*}
where $\tau_i^{(t)}=1 - u_i^{\prime(t)} v_i$.
\item
Update the parameter vector $\bm{\beta}_j$, $j=1,\ldots,m$, separately, by implementing Newton-Raphson method for one step. The first and second order derivatives of $Q$ with respect to $\bm{\beta}_j$ are given as follows:
\begin{eqnarray*}
\frac{\partial Q}{\partial \bm{\beta}_j} \hspace{-0.1in}&=&\hspace{-0.1in} 
\sum_{i=1}^{n}  \left[z_{ij} - \tau_i^{(t)} r_i^{(t)} \lambda_{ij} \right] \bm{x_{i}},\\
\frac{\partial^2 Q}{\partial \bm{\beta}_j \partial \bm{\beta}_j^\top}  
\hspace{-0.1in}&=&\hspace{-0.1in} 
-\sum_{i=1}^{n}  \tau_i^{(t)} r_i^{(t)}
\lambda_{ij} \bm{x_{i}}\bm{x_{i}}^\top.
\end{eqnarray*}

\item
Update the parameter $\phi$ by the following equation:
\begin{eqnarray*}
\phi^{(t+1)}=\phi^{(t)}-\frac{\sum_{i=1}^{n} \tau_i^{(t)} \left[\log \phi^{(t)} + 1 - \psi\left(\phi^{(t)}\right)+s_i^{(t)} - r_i^{(t)}\right]}{\left[ 1/\phi^{(t)} - \psi_1\left(\phi^{(t)} \right) \right] \sum_{i=1}^{n} \tau_i^{(t)}},
\end{eqnarray*}
where $\psi_1(\cdot)$ denotes the trigamma function.
\end{itemize}
\end{itemize}


\section{Multivariate zero-modified models}

\subsection{Definition}
To construct a multivariate zero-modified distribution, we first provide the definition for a multivariate zero-truncated distribution. Let $\bm{Y}=(Y_1, \ldots$, $Y_m)^\top$ denote a discrete random vector, where each $Y_j$ is defined on $\mathbb{N}$, then $\bm{W}'=(W'_1,\ldots,W'_m)^\top$ is said to follow a multivariate zero-truncated distribution if
\begin{eqnarray}
\bm{Y} \overset{d}{=} U_0\bm{W'}=\left\{ \begin{gathered}
  \bm{0_m}, \quad \hfill U_0=0,  \\
  \bm{W'}, \quad \hfill U_0=1,  \\
\end{gathered}  \right. \label{Model1}
\end{eqnarray}
where $U_0 \sim {\rm Bernoulli}(\pi_0)$, $\pi_0=\Pr(\bm{Y}\neq \bm{0})$, and $U_0$ is independent of $\bm{W}'$. The pmf of $\bm{W}'$ can be derived as
\begin{eqnarray}
\Pr(\bm{W}'=\bm{w}')=\frac{\Pr(\bm{Y}=\bm{w}')}{\pi_0},\quad \bm{w}' \neq \bm{0},
\end{eqnarray}
where $\bm{w}'=(w'_1,\ldots,w'_m)^\top$ is a vector of observed values. 

We now define our multivariate zero-modified distribution. $\bm{Z}=(Z_1,\ldots$, $Z_m)^\top$ is said to follow a multivariate zero-modified distribution if
\begin{eqnarray}
\bm{Z} \overset{d}{=} U'_0 \bm{W}' = \left\{ \begin{gathered}
  \bm{0}, \quad \hfill U'_0 = 0,  \\
  \bm{W}', \quad \hfill U'_0 = 1,  \\
\end{gathered}  \right. \label{Model2}
\end{eqnarray}
where $U'_0 \sim {\rm Bernoulli}(\pi'_0)$, $0<\pi'_0 < 1$, and $U'_0$ is independent of $\bm{W}'$. The pmf of $\bm{Z}$ can be derived as
\begin{eqnarray}
\Pr(\bm{Z}=\bm{z})=(1-\pi'_0)^{v}\left[\frac{\pi'_0}{\pi_0} \Pr(\bm{Y}=\bm{z})\right]^{1-v}, \label{MZM}
\end{eqnarray}
where $\bm{z}=(z_1,\ldots,z_m)^\top$ is a vector of observed values, $v=\mathbb{I}(\bm{z}=\bm{0})$. 

Several special cases of (\ref{MZM}) are given as follows.
\begin{itemize}
\item $0<\pi'_0<\pi_0$, $\bm{Z}$ follows a multivariate zero-inflated count distribution.
\item $\pi'_0=\pi_0$, $\bm{Z} \overset{d}{=} \bm{Y}$.
\item $\pi_0<\pi'_0<1$, $\bm{Z}$ follows a multivariate zero-deflated count distribution.
\end{itemize}

The covariates for $\pi'_0$ can be incorporated via a logit-link function:
\begin{eqnarray}
\pi'_0=\frac{\exp(\bm{x}^\top \bm{\gamma})}{1+\exp(\bm{x}^\top \bm{\gamma})}, 
\end{eqnarray}
where $\bm{x}=(1, x_{1},\ldots, x_{p})^\top$ and $\bm{\gamma}=(\gamma_{0},\gamma_{1},\ldots,\gamma_{p})^\top$.
\subsection{The models}

\subsubsection{Type I multivariate zero-modified Poisson model}
Let each $Y_j$ independently follow a Poisson distribution with parameter $\lambda_j$. Then $\bm{Z}$ is said to follow the Type I multivariate zero-modified Poisson (MZMP) distribution with parameters $\bm{\lambda}=(\lambda_1,\ldots, \lambda_m)^\top$ and  $\pi'_0$, denoted by $\bm{Z} \sim {\rm MZMP^{(I)}}(\bm{\lambda},\pi'_0)$. As a result, $\pi_0=1-e^{-\sum_{j=1}^m \lambda_j}$. The pmf of $\bm{Z}$ is
\begin{eqnarray}
\Pr(\bm{Z}=\bm{z})=(1-\pi'_0)^{v} 
\left(\frac{\pi'_0}{\pi_0}\prod_{j=1}^m \frac{\lambda_j^{z_j}e^{-\lambda_j}}{z_j!}\right)^{1-v},
\end{eqnarray}
where $v=\mathbb{I}(\bm{z}=\bm{0})$. The covariates can be incorporated via a log-link function:
\begin{eqnarray}
\lambda_{j}=\exp({\bm{x}^\top \bm{\beta}_j}), \quad j=1\ldots,m,
\end{eqnarray}
where $\bm{x}=(1, x_{1},\ldots, x_{p})^\top$ and $\bm{\beta}_j=(\beta_{j0},\beta_{j1},\ldots,\beta_{jp})^\top$.

\subsubsection{Type I multivariate zero-modified negative binomial model}

Let each $Y_j$ independently follow a negative binomial distribution with parameters $\lambda_j$ and $\phi_j$. Then $\bm{Z}$ is said to follow the Type I multivariate zero-modified negative binomial (MZMNB) distribution with parameters $\bm{\lambda}=(\lambda_1,\ldots, \lambda_m)^\top$, $\bm{\phi}=(\phi_1,\ldots, \phi_m)^\top$ and $\pi'_0$, denoted by $\bm{Z} \sim {\rm MZMNB^{(I)}}(\bm{\lambda},\bm{\phi},\pi'_0)$. As a result, $\pi_0=1-\prod_{j=1}^m \left(\frac{\phi_j}{\lambda_j+\phi_j} \right)^{\phi_j}$. The pmf of $\bm{Z}$ is
\begin{eqnarray}
\Pr(\bm{Z}=\bm{z}) \hspace{-0.1in} &=& \hspace{-0.1in} (1-\pi'_0)^{v} \left[\frac{\pi'_0}{\pi_0} \prod_{j=1}^m  \frac{\Gamma(z_j+\phi_j)}{\Gamma (\phi_j)z_j!}\left( \frac{\lambda_j} {\lambda_j+\phi_j} \right)^{z_j}\left(\frac{\phi_j}{\lambda_j +\phi_j} \right)^{\phi_j} \right]^{1-v},\nonumber \\
\end{eqnarray}
where $v=\mathbb{I}(\bm{z}=\bm{0})$. The covariates can be incorporated via a log-link function:
\begin{eqnarray}
\lambda_{j}=\exp({\bm{x}^\top \bm{\beta}_j}), \quad j=1\ldots,m,
\end{eqnarray}
where $\bm{x}=(1, x_{1},\ldots, x_{p})^\top$ and where $\bm{\beta}_j=(\beta_{j0},\beta_{j1},\ldots,\beta_{jp})^\top$.

\subsubsection{Type I multivariate zero-modified hurdle model}
Let each $Y_j$ independently follow a zero-modified distribution, which can be characterized as follows:
\begin{eqnarray}
Y_j \overset{d}{=} U_j W_j = \left\{ \begin{gathered}
  0, \quad \hfill U_j = 0,  \\
  W_j, \quad \hfill U_j = 1,  \\
\end{gathered}  \right. 
\end{eqnarray}
where $W_j$ follows a count distribution defined on $\mathbb{N}_{+}$, $U_j \sim Bernoulli(\pi_j)$, $0<\pi_j<1$, and $U_j$ is independent of $W_j$. Then $\bm{Z}$ is said to follow the Type I multivariate zero-modified hurdle (MZMH) distribution with parameters $\bm{\pi}=(\pi_1,\ldots, \pi_m)^\top$, $\bm{\Omega}=(\bm{\Omega}_1,\ldots, \bm{\Omega}_m)^\top$ and $\pi'_0$, denoted by $\bm{Z} \sim {\rm MZMH^{(I)}}(\bm{\pi},\bm{\Omega},\pi'_0)$. Here $\bm{\Omega}_j$ is the set of parameters related to $W_j$. As a result, $\pi_0=1-\prod_{j=1}^m(1-\pi_j)$. The pmf of $\bm{Z}$ is
\begin{eqnarray}
\Pr(\bm{Z}=\bm{z}) \hspace{-0.1in}&=&\hspace{-0.1in}
  (1-\pi'_0)^{v} \left[\frac{\pi'_0}{\pi_0} \prod_{j:z_j=0}(1-\pi_j) \prod_{j:z_j \ne 0}\pi_j f_{W_j}(z_j) \right]^{1-v},
\end{eqnarray}
The covariates for $\pi_j$ can be incorporated via a logit-link function:
\begin{eqnarray}
\pi_j=\frac{\exp(\bm{x}^\top \bm{\beta}_j)}{1+\exp(\bm{x}^\top \bm{\beta}_j)}, \quad j=1\ldots,m,
\end{eqnarray}
where $\bm{x}=(1, x_{1},\ldots, x_{p})^\top$ and where $\bm{\beta}_j=(\beta_{j0},\beta_{j1},\ldots,\beta_{jp})^\top$. The covariates for the location parameter $\lambda_j$ of $W_j$ can be incorporated via a log-link function:
\begin{eqnarray}
\lambda_{j}=\exp({\bm{x}^\top \bm{\alpha}_j}), \quad j=1\ldots,m,
\end{eqnarray}
where $\bm{x}=(1, x_{1},\ldots, x_{p})^\top$ and  $\bm{\alpha}_j=(\alpha_{j0},\alpha_{j1},\ldots,\alpha_{jp})^\top$.

\subsubsection{Type II multivariate zero-modified Poisson  model}
Let $\bm{Y}$ follow the multivariate Poisson distribution given in (\ref{MP}). Then $\bm{Z}$ is said to follow the Type II multivariate zero-modified Poisson (MZMP) distribution with parameters $\lambda_0$, $\bm{\lambda}=(\lambda_1,\ldots, \lambda_m)^\top$ and $\pi'_0$, denoted by $\bm{Z} \sim {\rm MZMP^{(II)}}(\lambda_0,\bm{\lambda},\pi'_0)$. As a result, $\pi_0=1-e^{-\sum_{j=0}^m \lambda_j}$. The pmf of $\bm{Z}$ is
\begin{eqnarray}
\Pr(\bm{Z}=\bm{z})=(1-\pi'_0)^{v} 
\left\{\frac{\pi'_0}{\pi_0}e^{-\lambda'}\sum_{n_0=0}^{l}\left[\frac{\lambda_0^{n_0}}{n_0!}
\prod_{j=1}^{m}\frac{\lambda_j^{z_j-n_0}}{(z_j-n_0)!}\right]\right\}^{1-v},
\end{eqnarray}
When $\lambda_0=0$, the Type II MZMP distribution simplifies to the Type I case. The covariates can be incorporated via a log-link function:
\begin{eqnarray}
\lambda_{j}=\exp({\bm{x}^\top \bm{\beta}_j}), \quad j=1\ldots,m,
\end{eqnarray}
where $\bm{x}=(1, x_{1},\ldots, x_{p})^\top$ and  $\bm{\beta}_j=(\beta_{j0},\beta_{j1},\ldots,\beta_{jp})^\top$. For ease of interpretation, we do not incorporate covariates in $\lambda_0$.

\subsubsection{Type II multivariate zero-modified negative binomial model}

Let $\bm{Y}$ follow the multivariate negative binomial distribution given in (\ref{MNB}). Then $\bm{Z}$ is said to follow the Type II multivariate zero-modified negative binomial (MZMNB) distribution with parameters $\bm{\lambda}=(\lambda_1,\ldots, \lambda_m)^\top$, $\phi$ and $\pi'_0$, denoted by $\bm{Z} \sim {\rm MZMNB^{(II)}}(\bm{\lambda},\phi,\pi'_0)$. As a result, $\pi_0=1- \left(\frac{\phi}{\lambda_{\cdot}+\phi} \right)^{\phi}$ where $\lambda_{\cdot}=\sum_{j=1}^{m} \lambda_j$. The pmf of $\bm{Z}$ is
\begin{eqnarray}
\Pr(\bm{Z}=\bm{z}) \hspace{-0.1in} &=& \hspace{-0.1in} (1-\pi'_0)^{v} \left[\frac{\pi'_0}{\pi_0}  \frac{\Gamma(\sum_{j=1}^{m}z_j+\phi)}{\Gamma(\phi) \prod_{j=1}^{m}z_j!}\frac{\phi^\phi \prod_{j=1}^{m}\lambda_j^{z_j}}{(\lambda_{\cdot}+\phi)^{\sum_{j=1}^{m}z_j+\phi}} \right]^{1-v},\nonumber \\
\end{eqnarray}
The covariates can be incorporated via a log-link function:
\begin{eqnarray}
\lambda_{j}=\exp({\bm{x}^\top \bm{\beta}_j}), \quad j=1\ldots,m,
\end{eqnarray}
where $\bm{x}=(1, x_{1},\ldots, x_{p})^\top$ and  $\bm{\beta}_j=(\beta_{j0},\beta_{j1},\ldots,\beta_{jp})^\top$.

\subsection{Inference}
Suppose we have a sample of size $n$. The corresponding observed values are $\bm{z}_1,\ldots, \bm{z}_n$, where $\bm{z}_i=(z_{i1},\ldots, z_{im})^\top$. The indicator variables are $v_1 ,\ldots, v_n$, where $v_i=\mathbb{I}(\bm{z}_i=\bm{0})$. Covariates are $\bm{x}_1,\ldots, \bm{x}_n$, where $\bm{x}_i=(1, x_{i1},\ldots, x_{ip})^\top$. The likelihood function then can be written as
\begin{eqnarray}
L(\bm{\gamma},\bm{\Theta}) \hspace{-0.1in}&=&\hspace{-0.1in} \prod_{i=1}^n (1-\pi'_{0i})^{v_i}
 \prod_{i=1}^n\left[\pi'_{0i}\frac{f_{\bm{Y}}(\bm{z}_i)}{1- f_{\bm{Y}}(\bm{0})} \right]^{1-v_i},
\end{eqnarray}
where $\pi'_{0i}=\frac{\exp(\bm{x}_i^\top \bm{\gamma})}{1+\exp(\bm{x}_i^\top \bm{\gamma})}$ and $\bm{\Theta}$ is the set of parameters related to $\bm{Y}$.

The observed log-likelihood function can be divided into two parts:
\begin{eqnarray*}
\ell_1(\bm{\gamma}) \hspace{-0.1in}&=&\hspace{-0.1in}
\sum_{i=1}^n v_i\log (1-\pi'_{0i})+
\sum_{i=1}^n (1-v_i)\log \pi'_{0i},\\
\ell_2(\bm{\Theta}) \hspace{-0.1in}&=&\hspace{-0.1in}\sum_{i=1}^n (1-v_i) \left\{\log f_{\bm{Y}}(\bm{z}_i)- 
\log\left[1- f_{\bm{Y}}(\bm{0})\right]\right\}\\
\hspace{-0.1in}&=&\hspace{-0.1in}
\sum_{i \in \mathbb{I}}\left\{\log f_{\bm{Y}}(\bm{z}_i)- 
\log\left[1- f_{\bm{Y}}(\bm{0})\right]\right\},
\end{eqnarray*}
where $\mathbb{I}=\{i|v_i=0, i=1,\ldots,n\}$. Thus, the maximization procedure can be completed for $\ell_1$ and $\ell_2$ respectively. For $\ell_1$, the estimates for $\bm{\gamma}$ can be obtained through logistic regression. For $\ell_2$, we use the MM algorithm to maximize it.

As shown in \cite{Zhou-Lange-2010}, we have
\begin{eqnarray}
-\log(1-\alpha) \geq -\log(1-\alpha_0)+\frac{\alpha_0}{1-\alpha_0} \log \left(\frac{\alpha}{\alpha_0}\right).
\label{MM_inequality}
\end{eqnarray}
In $\ell_2$, we apply \eqref{MM_inequality} with $\alpha = f_{\bm{Y}}(\bm{0}|\bm{\Theta})$ and  $\alpha_0 = f_{\bm{Y}}\left(\bm{0}|\bm{\Theta}^{(t)}\right)$, and obtain
\begin{eqnarray*}
\ell_2(\bm{\Theta}) 
\hspace{-0.1in}&\geq&\hspace{-0.1in} 
\sum_{i \in \mathbb{I}} \left\{\log f_{\bm{Y}} (\bm{z}_i|\bm{\Theta}) -  
\log\left[1 - f_{\bm{Y}}\left(\bm{0}|\bm{\Theta}^{(t)}\right)\right] + \frac{f_{\bm{Y}}\left(\bm{0}|\bm{\Theta}^{(t)}\right)}{1-f_{\bm{Y}}\left(\bm{0}|\bm{\Theta}^{(t)}\right)} \log \left[\frac{f_{\bm{Y}}(\bm{0}|\bm{\Theta})}{f_{\bm{Y}}\left(\bm{0}|\bm{\Theta}^{(t)}\right)}\right] \right\} \\
\hspace{-0.1in}&=&\hspace{-0.1in} 
\sum_{i \in \mathbb{I}} \left\{\log f_{\bm{Y}} (\bm{z}_i|\bm{\Theta}) + \frac{f_{\bm{Y}}\left(\bm{0}|\bm{\Theta}^{(t)}\right)}{1-f_{\bm{Y}}\left(\bm{0}|\bm{\Theta}^{(t)}\right)} \log f_{\bm{Y}}(\bm{0}|\bm{\Theta}) \right\} + C \\
\hspace{-0.1in}&\triangleq&\hspace{-0.1in} 
Q(\bm{\Theta}; \bm{\Theta}^{(t)}),
\end{eqnarray*}
where $C$ is a constant not related to the parameter vector $\bm{\Theta}$, and $Q$ is the surrogate function. 

\subsubsection{Case 1: Type I MZMP model}
\begin{itemize}
\item
Minorization:
The surrogate function of $\ell_2$ can be written as
\begin{eqnarray*}
Q\left(\bm{\Theta}; \bm{\Theta}^{(t)}\right) \propto \sum_{j=1}^m \sum_{i \in \mathbb{I}} \left(z_{ij} \log\lambda_{ij} - u_i^{\prime(t)} \lambda_{ij} \right),
\end{eqnarray*}
where $u_i^{\prime(t)}=1/u_i^{(t)}$ with $u_i^{(t)}=1-e^{-\sum_{j=1}^m \lambda_{ij}^{(t)}}$.

\item
Maximization:
Update the parameter vector $\bm{\beta}_j$, $j=1,\ldots,m$, separately, by implementing the Newton-Raphson method for one step. The first and second order derivatives of $Q$ with respect to $\bm{\beta}_j$ are given as follows:
\begin{eqnarray*}
\frac{\partial Q}{\partial \bm{\beta}_j} \hspace{-0.1in}&=&\hspace{-0.1in} 
\sum_{i \in \mathbb{I}}  \left( z_{ij}- u_i^{\prime(t)}\lambda_{ij}\right) \bm{x_{i}},\\
\frac{\partial^2 Q}{\partial \bm{\beta}_j\partial \bm{\beta}_j^\top} \hspace{-0.1in}&=&\hspace{-0.1in} -\sum_{i \in \mathbb{I}}  u_i^{\prime(t)} \lambda_{ij} \bm{x_{i}}\bm{x_{i}}^\top.
\end{eqnarray*}

\end{itemize}

\subsubsection{Case 2: Type I MZMNB model}
\begin{itemize}
\item
Minorization:
The surrogate function of $\ell_2$ can be written as
\begin{eqnarray*}
Q\left(\bm{\Theta}; \bm{\Theta}^{(t)}\right) 
\hspace{-0.1in}&\propto&\hspace{-0.1in}
\sum_{j=1}^m \sum_{i \in \mathbb{I}} \left[ \log\left(\Gamma(z_{ij}+\phi_j) \right) -
\log \left(\Gamma(\phi_j) \right) + z_{ij} \log \left(\frac{\lambda_{ij}}{\lambda_{ij}+\phi_j}\right) \right.  \\
&&\hspace{-0.1in} \left.
+u_i^{\prime(t)} \phi_j \log \left( \frac{\phi_j}{\lambda_{ij}+\phi_j} \right) \right],
\end{eqnarray*}
where $u_i^{\prime(t)}=1/u_i^{(t)}$ with $u_i^{(t)}=1-\prod_{j=1}^m \left( \frac{\phi_j^{(t)}}{\lambda_{ij}^{(t)}+\phi_j^{(t)}} \right)^{\phi_j^{(t)}}$.

\item
Maximization:
Update the parameters $\bm{\beta}_j$ and $\phi_j$, $j=1,\ldots,m$, for each margin by implementing the Newton-Raphson method for one step. The first and second order derivatives of $Q$ with respect to $\bm{\beta}_j$ and $\phi_j$ are given as follows:
\begin{eqnarray*}
\frac{\partial Q}{\partial \bm{\beta}_j} \hspace{-0.1in}&=&\hspace{-0.1in}
\sum_{i \in \mathbb{I}} \frac{\left( z_{ij }- u_i^{\prime(t)} \lambda_{ij} \right) \phi _j}{\lambda_{ij}+\phi_j}\bm{x_i},  \\
\frac{\partial Q}{\partial \phi_j} \hspace{-0.1in}&=&\hspace{-0.1in}
\sum_{i \in \mathbb{I}} \left[ \psi(z_{ij}+\phi_j) - \psi(\phi_j) + u_i^{\prime(t)} \log\frac{\phi_j}{\lambda_{ij}+\phi_j} + \frac{u_i^{\prime(t)} \lambda_{ij}-  z_{ij}} {\lambda_{ij}+\phi_j}\right],\\
\frac{\partial^2 Q}{\partial \bm{\beta}_j \partial \bm{\beta}_j^\top} 
\hspace{-0.1in}&=&\hspace{-0.1in}  
-\sum_{i \in \mathbb{I}}  \frac{ \left( z_{ij} + u_i^{\prime(t)} \phi_j \right) \lambda_{ij} \phi_j} {(\lambda_{ij}+\phi_j)^2} \bm{x_i}\bm{x_i}^\top, \\
\frac{\partial^2 Q}{\partial \phi_j^2} \hspace{-0.1in}&=&\hspace{-0.1in}
\sum_{i \in \mathbb{I}} \left[ \psi_1(z_{ij}+\phi_j) - \psi_1(\phi_j) + \frac{ u_i^{\prime(t)} \lambda_{ij}^2 +  z_{ij} \phi _j}{\phi_j(\lambda_{ij}+\phi_j)^2} \right], \\
\frac{\partial^2 Q}{\partial \bm{\beta}_j\partial \phi_j} \hspace{-0.1in}&=&\hspace{-0.1in}   
\sum_{i \in \mathbb{I}} \frac{\left ( z_{ij} - u_i^{\prime(t)} \lambda_{ij} \right) \lambda_{ij}}{(\lambda_{ij} + \phi_j)^2} \bm{x_i}.
\end{eqnarray*}

\end{itemize}

\subsubsection{Case 3: Type I MZMH model}
Denote $\bm{Z'}=(Z'_1,\ldots,Z'_m)^\top$ where $Z'_j=\mathbb{I}(Z_j>0)$. The corresponding observed values are denoted by $\bm{z}'_1,\ldots,\bm{z}'_n$ where $\bm{z}'_i=(z'_{i1},\ldots,z'_{im})^\top$. As pointed out in \cite{Zhang-Calderin-Li-Wu-2020}, $\ell_2(\bm{\Theta})$ can be further decomposed into two parts,
\begin{eqnarray*}
\ell_2^{(1)}(\bm{\Theta}_1)
\hspace{-0.1in}&=&\hspace{-0.1in}
\sum_{i \in \mathbb{I}} \sum_{j=1}^m  \left[z'_{ij}\log
\pi_{ij}+(1-z'_{ij})\log (1-\pi_{ij}) \right]-\sum_{i \in \mathbb{I}} \log \left[1-\prod_{j=1}^m (1-\pi_{ij}) \right],\\
\ell_2^{(2)}(\bm{\Theta}_2) 
\hspace{-0.1in}&=&\hspace{-0.1in} 
\sum_{i \in \mathbb{I}} \sum_{j=1}^m  z'_{ij} \log f_{W'_{j}}(z_{ij}),
\end{eqnarray*}
where $\bm{\Theta}_1=\bm{\beta}$ is the parameter set linked to all $\pi_{j}$, and $\bm{\Theta}_2$ is the parameter set related to all $W_{j}$.

Thus, the maximization procedure can be completed for $\ell_2^{(1)}$ and  $\ell_2^{(2)}$ respectively. For $\ell_2^{(2)}$, the estimation can proceed in respect of the zero-truncation part of each margin separately. For $\ell_2^{(1)}$, we implement the MM algorithm as described below. 

\begin{itemize}
\item
Minorization:
The surrogate function of $\ell_2^{(1)}$ can be written as
\begin{eqnarray*}
Q\left(\bm{\beta}; \bm{\beta}^{(t)}\right) 
\hspace{-0.1in}&\propto&\hspace{-0.1in}
\sum_{j=1}^m \sum_{i \in \mathbb{I}}  \left[{z'_{ij}} \log {\pi_{ij}} + \left( u_i^{\prime(t)} - z'_{ij} \right)
\log (1-{\pi_{ij}}) \right] ,
\end{eqnarray*}
where $u_i^{\prime(t)}=1/u_i^{(t)}$ with $u_i^{(t)}=1-\prod_{j=1}^m \left(1-\pi_{ij}^{(t)}\right)$.

\item
Maximization:
Update the parameter vector $\bm{\beta}_j$, $j=1,\ldots,m$, separately, by implementing the Newton-Raphson method for one step. The first and second order derivatives of $Q$ with respect to $\bm{\beta}_j$ are given as follows:
\begin{eqnarray*}
\frac{\partial Q}{\partial \bm{\beta}_j} \hspace{-0.1in}&=&\hspace{-0.1in} 
\sum_{i \in \mathbb{I}} \left( z'_{ij} - u_i^{\prime(t)}  \pi_{ij} \right)\bm{x}_i, \\
\frac{\partial^2 Q}{\partial \bm{\beta}_j\partial \bm{\beta}_j^\top} 
\hspace{-0.1in}&=&\hspace{-0.1in} 
-\sum_{i \in \mathbb{I}} u_i^{\prime(t)} {\pi_{ij}(1-\pi_{ij})}\bm{x}_i \bm{x}_i^\top.
\end{eqnarray*}
\end{itemize}

\subsubsection{Case 4: Type II MZMP model}
Due to the complicated form of $\sum_{i \in \mathbb{I}} \log f_{\bm{Y}}(\bm{z}_i|\bm{\Theta})$, we need to further find a surrogate function of it. Obtained from the relationship between the EM and MM algorithms as mentioned in the introduction, the surrogate function of this term can be written as 
\begin{eqnarray*}
Q_1\left(\bm{\Theta}; \bm{\Theta}^{(t)}\right) \hspace{-0.1in}&\propto&\hspace{-0.1in} 
\sum_{i \in \mathbb{I}} \log \left[ 
e^{-\lambda'} \frac{\lambda_0^{n_{0i}^{(t)}}}{n_{0i}^{(t)}!}
\prod_{j=1}^{m}\frac{\lambda_j^{y_j-n_{0i}^{(t)}}}{\left(y_j-n_{0i}^{(t)}\right)!} \right],
\end{eqnarray*}
where 
\begin{eqnarray*}
n_{0i}^{(t)} =
\left\{ \begin{gathered}
 \frac{\lambda_0^{(t)} f_{\bm{Y}}\left(\bm{z}_i-\bm{1} | \bm{\Theta}^{(t)}\right)}
{f_{\bm{Y}}\left(\bm{z}_i | \bm{\Theta}^{(t)}\right)}, \quad \hfill \min(z_{i1}, \ldots,z_{im}) > 0,  \\
  0, \quad \hfill\min(z_{i1}, \ldots,z_{im}) = 0,  \\
\end{gathered}  \right.
\end{eqnarray*}
and $\bm{1}=(1,\ldots,1)^\top$ denotes the vector of dimension $m$ with all elements equal to 1.

\begin{itemize}
\item
Minorization:
The surrogate function of $\ell_2$ then can be written as
\begin{eqnarray*}
Q\left(\bm{\Theta}; \bm{\Theta}^{(t)}\right) \hspace{-0.1in}&\propto&\hspace{-0.1in} 
\sum_{i \in \mathbb{I}} 
\left(n_{0i}^{(t)} \log \lambda_0 - u_i^{\prime(t)} \lambda_0 \right)\\
&&\hspace{-0.1in}
+\sum_{i \in \mathbb{I}} \sum_{j=1}^{m}   \left[\left(z_{ij}-n_{0i}^{(t)}\right) \log \lambda_{ij} - u_i^{\prime(t)} \lambda_{ij} \right],
\end{eqnarray*}
where $u_i^{\prime(t)}=1/u_i^{(t)}$ with $u_i^{(t)}=1-e^{-\lambda_0^{(t)}-\sum_{j=1}^m \lambda_{ij}^{(t)}}$.

\item
Maximization:
\begin{itemize}
\item
Update the parameter $\lambda_0$ by the following equation:
\begin{eqnarray*}
\lambda_0^{(t+1)}=\frac{\sum_{i \in \mathbb{I}} n_{0i}^{(t)}}{\sum_{i \in \mathbb{I}} u_i^{\prime(t)}}.
\end{eqnarray*}
\item
Update the parameter vector $\bm{\beta}_j$, $j=1,\ldots,m$, separately, by implementing the Newton-Raphson method for one step. The first and second order derivatives of $Q$ with respect to $\bm{\beta}_j$ are given as follows:
\begin{eqnarray*}
\frac{\partial Q}{\partial \bm{\beta}_j} \hspace{-0.1in}&=&\hspace{-0.1in} 
\sum_{i \in \mathbb{I}}  \left( z_{ij} - n_{0i}^{(t)} - u_i^{\prime(t)}\lambda_{ij}^{(t)} \right) \bm{x_{i}},\\
\frac{\partial^2 Q}{\partial \bm{\beta}_j\partial \bm{\beta}_j^\top} 
\hspace{-0.1in}&=&\hspace{-0.1in} -\sum_{i \in \mathbb{I}}  u_i^{\prime(t)} \lambda_{ij}^{(t)} \bm{x_{i}}\bm{x_{i}}^\top.
\end{eqnarray*}
\end{itemize}

\end{itemize}

\subsubsection{Case 5: Type II MZMNB model}
Due to the complicated form of $\sum_{i \in \mathbb{I}} \log f_{\bm{Y}}(\bm{z}_i|\bm{\Theta})$ and $ \sum_{i \in \mathbb{I}} \log f_{\bm{Y}}(\bm{0}|\bm{\Theta})$, we need to further find  surrogate functions of these two terms. Obtained from the relationship between the EM and MM algorithms, the surrogate function of the first term can be written as 
\begin{eqnarray*}
Q_1\left(\bm{\Theta}; \bm{\Theta}^{(t)}\right) \hspace{-0.1in}&\propto&\hspace{-0.1in} 
\sum_{i \in \mathbb{I}} \sum_{j=1}^m \left[ z_{ij} \log \lambda_{ij} -r_{1i}^{(t)} \lambda_{ij} \right]\\
&&\hspace{-0.1in}
+\sum_{i \in \mathbb{I}}  \left[ \phi \log \phi - \log\Gamma(\phi) + (\phi-1) s_{1i}^{(t)} - \phi r_{1i}^{(t)} \right],
\end{eqnarray*}
where 
\begin{eqnarray*}
r_{1i}^{(t)}
\hspace{-0.1in} &=& \hspace{-0.1in} 
\frac{ \sum_{j=1}^{m} z_{ij}+\phi^{(t)}} {\sum_{j=1}^{m}\lambda_{ij}^{(t)}+\phi^{(t)}},\\
s_{1i}^{(t)} 
\hspace{-0.1in} &=& \hspace{-0.1in} 
\psi\left(\sum_{j=1}^{m} z_{ij} + \phi^{(t)} \right) - \log \left (\sum_{j=1}^{m}\lambda_{ij}^{(t)} + \phi^{(t)}\right),
\end{eqnarray*}
and the surrogate function of the second term can be written as 
\begin{eqnarray*}
Q_2\left(\bm{\Theta}; \bm{\Theta}^{(t)}\right) \hspace{-0.1in}&\propto&\hspace{-0.1in} 
-\sum_{i \in \mathbb{I}} \sum_{j=1}^m r_{2i}^{(t)} \lambda_{ij}
+\sum_{i \in \mathbb{I}}  \left[ \phi \log \phi - \log\Gamma(\phi) + (\phi-1) s_{2i}^{(t)} - \phi r_{2i}^{(t)} \right],
\end{eqnarray*}
where 
\begin{eqnarray*}
r_{2i}^{(t)}
\hspace{-0.1in} &=& \hspace{-0.1in} 
\frac{\phi^{(t)}} {\sum_{j=1}^{m}\lambda_{ij}^{(t)}+\phi^{(t)}},\\
s_{2i}^{(t)} 
\hspace{-0.1in} &=& \hspace{-0.1in} 
\psi\left( \phi^{(t)} \right) - \log \left (\sum_{j=1}^{m}\lambda_{ij}^{(t)} + \phi^{(t)}\right).
\end{eqnarray*}

\begin{itemize}
\item
Minorization:
The surrogate function of $\ell_2$ then can be written as
\begin{eqnarray*}
Q\left(\bm{\Theta}; \bm{\Theta}^{(t)}\right) 
\hspace{-0.1in} &\propto& \hspace{-0.1in} 
\sum_{j=1}^m \sum_{i \in \mathbb{I}} \left( z_{ij} \log\lambda_{ij} - r_i^{\prime(t)}  \lambda_{ij} \right)\\
&&\hspace{-0.1in}
+\sum_{i \in \mathbb{I}} \left[ u_i^{\prime(t)} \phi \log \phi -  u_i^{\prime(t)} \log\Gamma(\phi) + \left( s_i^{\prime(t)} - r_i^{\prime(t)} \right) \phi \right],
\end{eqnarray*}
where $u_i^{\prime(t)}=1/u_i^{(t)}$ with $u_i^{(t)}=1- \left(\frac{\phi^{(t)}}{\sum_{j=1}^m\lambda_{ij}^{(t)}+\phi^{(t)}} \right)^{\phi^{(t)}}$, $r_i^{\prime(t)}=r_{1i}^{(t)} + \left( u_i^{\prime(t)} - 1 \right)r_{2i}^{(t)}$ and  $s_i^{\prime(t)}=s_{1i}^{(t)} + \left( u_i^{\prime(t)} - 1 \right)s_{2i}^{(t)}$.

\item
Maximization:
\begin{itemize}
\item
Update the parameter vector $\bm{\beta}_j$, $j=1,\ldots,m$, separately, by implementing the Newton-Raphson method for one step. The first and second order derivatives of $Q$ with respect to $\bm{\beta}_j$ are given as follows:
\begin{eqnarray*}
\frac{\partial Q}{\partial \bm{\beta}_j} \hspace{-0.1in}&=&\hspace{-0.1in} 
\sum_{i \in \mathbb{I}}  \left( z_{ij} - r_i^{\prime(t)} \lambda_{ij} \right) \bm{x_{i}},\\
\frac{\partial^2 Q}{\partial \bm{\beta}_j\partial \bm{\beta}_j^\top} 
\hspace{-0.1in}&=&\hspace{-0.1in} 
-\sum_{i \in \mathbb{I}}  r_i^{\prime(t)} \lambda_{ij} \bm{x_{i}}\bm{x_{i}}^\top.
\end{eqnarray*}
\item
Update the parameter $\phi$ by the following equation:
\begin{eqnarray*}
\phi^{(t+1)} = \phi^{(t)} - \frac{  \sum_{i \in \mathbb{I}} \left\{ u_i^{\prime(t)} \left[\log \phi^{(t)} + 1 - \psi\left(\phi^{(t)}\right) \right]+  s_i^{\prime(t)}-r_i^{\prime(t)} \right\}}{ \left[ 1/\phi^{(t)}-\psi_1 \left(\phi^{(t)}\right) \right] \sum_{i \in \mathbb{I}} u_i^{\prime(t)} }.
\end{eqnarray*}
\end{itemize}

\end{itemize}

\section{Application}
\subsection{Data description}
This application is based on an automobile portfolio from a major insurance company operating in Spain in 1995. The dataset contains information for 80,994 policyholders. Eleven covariates are considered in our analysis. The detailed description for each predictor is presented in Table \ref{covariates}. The mean of each covariate is also provided in the table to show the proportion of the corresponding group. For example, the mean of $v1$ tells us that 16.0\%  of policyholders are female. The simplest policy only includes third-party liability (denoted as $Z_1$ type) and a set of basic guarantees such as emergency roadside assistance, legal assistance or insurance covering medical costs (denoted as $Z_2$ type). The comprehensive coverage (damage to one’s vehicle caused by any unknown party, for example, damage resulting from theft, flood or fire) and the collision coverage  (damage resulting from a collision with another vehicle or object when the policyholder is at fault) are excluded from this simplest policy. This simplest type of policy forms the baseline group, while the variable $v9$ denotes the policies which also include comprehensive coverage (except fire), and the variable $v10$ denotes policies which also include comprehensive and collision coverage. The empirical joint distribution for claim numbers $Z_1$ and $Z_2$ is displayed in Table \ref{distribution}. The overall Pearson's correlation coefficient between these two types of claim is 0.187. This dataset was previously used in \cite{Bermudez-2009} and \cite{Bermudez-Karlis-2012} for analysis of bivariate count models.

\begin{table}[htbp]
\setlength{\belowcaptionskip}{10pt}
\caption{The description for explanatory variables.}
\centering
\begin{tabular}{llc}
\toprule
\multicolumn {1}{c}{Variable} & \multicolumn {1}{c}{Description} & Mean \\
\midrule
$v1$ & = 1 for women; = 0 for men & 0.160\\
$v2$ & = 1 when driving in urban area; = 0 otherwise & 0.669\\
$v3$ & = 1 when zone is medium risk (Madrid and Catalonia) & 0.239\\
$v4$ & = 1 when zone is high risk (northern Spain) & 0.194\\
$v5$ & = 1 if the driving license is between 4 and 14 years old & 0.257\\
$v6$ & = 1 if the driving license is 15 or more years old & 0.719\\
$v7$ & = 1 if the client is in the company for more than 5 years & 0.856\\
$v8$ & = 1 if the insured is 30 years old or younger & 0.092\\
$v9$ & = 1 if includes comprehensive coverage (except fire) & 0.156\\
$v10$ & = 1 if includes comprehensive and collision coverage & 0.353\\
$v11$ & = 1 if horsepower is $\geq$ 5,500 cc & 0.806\\
\bottomrule
\end{tabular}
\label{covariates}
\end{table}

\begin{table}[htbp]
\setlength{\belowcaptionskip}{10pt}
\caption{The empirical joint distribution of $Z_1$ and $Z_2$.}
\centering
\begin{tabular}{crrrrrrrr}
\toprule
\multicolumn {1}{c}{$Z_1$} & \multicolumn {8}{c}{$Z_2$}\\
\cmidrule(lr) {2-9}
  & 0 & 1 & 2 & 3 & 4 & 5 & 6 & 7 \\
\midrule
0 & 71,087 & 3,722 & 807 & 219 & 51 & 14 & 4 & 0\\
1 &  3,022 &   686 & 184 &  71 & 26 & 10 & 3 & 1\\
2 &    574 &   138 &  55 &  15 &  8 &  4 & 1 & 1\\
3 &    149 &    42 &  21 &   6 &  6 &  1 & 0 & 1\\
4 &     29 &    15 &   3 &   2 &  1 &  1 & 0 & 0\\
5 &      4 &     1 &   0 &   0 &  0 &  0 & 2 & 0\\
6 &      2 &     1 &   0 &   1 &  0 &  0 & 0 & 0\\
7 &      1 &     0 &   0 &   1 &  0 &  0 & 0 & 0\\
8 &      0 &     0 &   1 &   0 &  0 &  0 & 0 & 0\\
\bottomrule
\end{tabular}
\label{distribution}
\end{table}

\subsection{Model comparison}
We start with the case when no covariates are introduced. Altogether, 15 models are fitted and compared. The first five models are standard multivariate count models without incorporating zero-inflation or zero-modification. Among these five models, the first three are fitted with independent margins. We consider the multivariate independent Poisson (MIP),  the multivariate independent negative binomial (MNB), and the multivariate independent hurdle (MIH) as our potential choices. In the subsequent two cases, dependence among margins is considered. Two models are fitted, namely the multivariate Poisson (MP) and the multivariate negative binomial (MNB). The middle five models are the corresponding multivariate zero-inflated versions of the first five models, and the last five models are the corresponding multivariate zero-modified versions. It is worth mentioning that for the three relevant hurdle models (MIH, Type I MZIH, Type I MZMH), the distributions for the two univariate zero-truncated parts $W_1$ and $W_2$ are both chosen as unit-shifted negative binomial (USNB). We make this choice according to the $\chi^2$ statistics and log-likelihood values as shown in Table \ref{goodness}. The overall comparison of the 15 models is displayed in Table \ref{comparison1}. 

\begin{table}[htbp]
\setlength{\belowcaptionskip}{10pt}
\caption{Goodness-of-fit of marginal models.}
\centering
\begin{tabular}{crrrrr}
\toprule
$W_1$ & \multicolumn{1}{c}{Observed} & \multicolumn{1}{c}{ZTP} & \multicolumn{1}{c}{ZTNB} & \multicolumn{1}{c}{USP} & \multicolumn{1}{c}{USNB}  \\
\midrule
1 & 4,003 & 3,859.22 & 4,026.70 & 3,814.73 & 3,999.98 \\
2 &   796 & 1,023.18 &   780.45 & 1,100.20 &   813.68 \\
3 &   226 &   180.85 &   199.72 &   158.65 &   202.65 \\
4 &    51 &    23.97 &    57.31 &    15.25 &    53.55 \\
5 &     7 &     2.54 &    17.51 &     1.10 &    14.56 \\
$\geq$6&7 &     0.24 &     8.31 &     0.07 &     5.59 \\
\midrule
$\chi^2$& &   293.32 &    11.12 &   958.45 &     7.48 \\
LogLik &  &-3,546.53 &-3,483.17 &-3,604.39 &-3,481.01 \\
\midrule
$W_2$ & \multicolumn{1}{c}{Observed} & \multicolumn{1}{c}{ZTP} & \multicolumn{1}{c}{ZTNB} & \multicolumn{1}{c}{USP} & \multicolumn{1}{c}{USNB}  \\
\midrule
1 &  4,605 & 4,375.24 & 4,649.53 & 4,302.22 & 4,603.02 \\
2 &  1,071 & 1,398.38 & 1,026.10 & 1,520.45 & 1,079.10 \\
3 &    315 &   297.96 &   298.98 &   268.67 &   308.13 \\
4 &     92 &    47.62 &    97.68 &    31.65 &    93.24 \\
5 &     30 &     6.09 &    33.99 &     2.80 &    29.01 \\
$\geq$6&13 &     0.71 &    19.73 &     0.21 &    13.50 \\
\midrule
$\chi^2$&  &   436.79 &     6.34 & 1,321,19 &     0.28 \\
LogLik &   &-4,864.86 &-4,755.15 &-4,963.00 &-4,751.31 \\
\bottomrule
\end{tabular}
\label{goodness}
\end{table}

\begin{table}[htbp]
\setlength{\belowcaptionskip}{10pt}
\caption{Comparison results of 15 fitted models without covariates incorporated.}
\centering
\begin{tabular}{ccccc}
\toprule
Model & Parameters & LogLik & AIC & BIC\\
\midrule
MIP  & 2 & -53,271.05 & 106,546.10 & 106,564.70\\
MINB & 4 & -48,949.67 &  97,907.34 &  97,944.55\\
MIH  & 6 & -48,948.02 &  97,908.03 &  97,963.85\\
MP   & 3 & -52,283.93 & 104,573.90 & 104,601.80\\
MNB  & 3 & -48,314.53 &  96,635.06 &  96,662.97\\
\midrule
Type I MZIP   & 3 & -48,630.52 & 97,267.03 & 97,294.94\\
Type I MZINB  & 5 & -48,101.02 & 96,212.03 & 96,258.54\\
Type I MZIH   & 7 & -48,087.96 & 96,189.91 & 96,255.03\\
Type II MZIP  & 4 & -48,630.52 & 97,269.03 & 97,306.24\\
Type II MZINB & 4 & -48,310.44 & 96,628.88 & 96,666.09\\
\midrule
Type I MZMP   & 3 & -48,630.52 & 97,267.03 & 97,294.94\\
Type I MZMNB  & 5 & -48,101.02 & 96,212.03 & 96,258.54\\
Type I MZMH   & 7 & -48,087.96 & 96,189.91 & 96,255.03 \\
Type II MZMP  & 4 & -48,630.52 & 97,269.03 & 97,306.24 \\
Type II MZMNB & 4 & -48,310.44 & 96,628.88 & 96,666.09 \\
\bottomrule
\end{tabular}
\label{comparison1}
\end{table}

Several observations can be made from the table. First, the five multivariate zero-inflated models significantly outperform their counterparts in the first group in terms of the information criteria. This reveals the multivariate zero-inflation feature exhibited in the dataset. Second, the five multivariate zero-inflated models are equivalent to their corresponding multivariate zero-modified ones when no covariates are introduced. The only difference between each pair lies in the parameterization. Third, by comparing the MP model with the MIP model and the MNB model with the MINB model in the first group, we observe that the models considering dependence outperform the independent models. However, after considering the dependence resulting from extra common zeros, the two multivariate zero-inflated/zero-modified models derived from MP and MNB are inferior to those derived from MIP and MINB. Specifically, the dependence parameter $\lambda_0$ in the Type II MZIP/MZMP model is approximately 0, which supports the better performance of the Type I MZIP/MZMP model with one fewer parameter involved. We also observe substantial improvements from the Type II MZINB/MZMNB model to the Type I MZINB/MZMNB model. Finally, all the performance measures suggest the superiority of the Type I MZIH/MZMH model. The separation of the positive part from the zero part in the hurdle model gives it plenty of freedom to deal with various features reflected in the margins. 

\subsection{Model fitting}
We next focus on the comparison when covariates are introduced in these multivariate zero-inflated and zero-modified models. The corresponding results are presented in Table \ref{comparison2}. It is worth mentioning that for the zero-truncated parts $W_1$ and $W_2$ in the Type I MZIH/MZMH models, only significant predictors are retained in order to avoid the overfitting problem. As seen from the table, the five multivariate zero-inflated models uniformly perform better than their corresponding multivariate zero-modified versions in this case. However, the difference between each pair is not that substantial. Different from what we have seen in Table \ref{comparison1}, the Type II MZINB and Type II MZMNB models now work better than the Type I MZINB and Type I MZMNB models. Again, the Type I MZIH and Type I MZMH models outperform their alternatives in each group.

\begin{table}[htbp]
\setlength{\belowcaptionskip}{10pt}
\caption{Comparison results of 10 fitted models with covariates incorporated.}
\centering
\begin{tabular}{ccccc}
\toprule
Model & Parameters & LogLik & AIC & BIC\\
\midrule
Type I MZIP   & 36 & -45,171.71 & 90,415.42 & 90,750.30 \\
Type I MZINB  & 38 & -44,965.58 & 90,007.16 & 90,360.64 \\
Type I MZIH   & 41 & -44,704.11 & 89,490.22 & 89,871.61 \\
Type II MZIP  & 37 & -45,167.26 & 90,408.51 & 90,752.69 \\
Type II MZINB & 37 & -44,914.28 & 89,902.56 & 90,246.74 \\
\midrule
Type I MZMP   & 36 & -45,180.93 & 90,433.87 & 90,768.74 \\
Type I MZMNB  & 38 & -44,975.14 & 90,026.27 & 90,379.75 \\
Type I MZMH   & 41 & -44,718.68 & 89,519.36 & 89,900.75 \\
Type II MZMP  & 37 & -45,176.60 & 90,427.20 & 90,771.37 \\
Type II MZMNB & 37 & -44,922.15 & 89,918.31 & 90,262.49 \\
\bottomrule
\end{tabular}
\label{comparison2}
\end{table}

We now turn attention to the parameter estimates in our models. For brevity, we only report the estimation results regarding the Type I MZIH model and the Type I MZMH model in Table \ref{MZIH1} and Table \ref{MZTH1}, respectively. Focusing on the Type I MZIH model, we notice that $v1$, $v3$-$v4$, $v7$-$v10$ are all statistically significant in modeling the multivariate zero-inflation parameter $\pi_0$. For $\pi_1$, $v4$, $v9$-$v10$ are significant predictors. It can be concluded that  policies with comprehensive and collision coverage ($v9$, $v10$) are associated with decreased chances of a claim in this category. We also note that driving in a high-risk region ($v4$) could increase the chances of making a claim of this type. For $\pi_2$, $v5$, $v7$ and $v9$-$v11$ are significant predictors. This tells us that holding a driver's license between 4 and 14 years ($v5$), and that policies with comprehensive and collision coverage ($v9$, $v10$), and cars with greater horsepower ($v11$) are all associated with an increase the occurrence probability for $Z_2$ type. However, policyholders with the company for more than 5 years ($v7$) exhibit a lower probability of claiming for this type. For the zero-truncated parts $W_1$ and $W_2$, only $v9$ is a significant predictor for the expected number of $W_2$. Concentrating on the Type I MZMH model, $v3$ and $v7$-$v11$ are statistically associated with the multivariate zero-modification parameter $\pi'_0$. Similar conclusions can be made regarding the significance of parameters embedded in $\pi_1$ and $\pi_2$. The model estimates for the two zero-truncated parts are same as in the Type I MZIH model.

\begin{table}[htbp]
\setlength{\belowcaptionskip}{10pt}
\caption{Estimation results under the Type I MZIH model.}
\centering
\begin{tabular}{lr@{.}lr@{.}lr@{.}lr@{.}lr@{.}lr@{.}l}
\toprule
& \multicolumn{4}{c}{$\pi_0$} & \multicolumn{4}{c}{$\pi_1$} 
& \multicolumn{4}{c}{$\pi_2$}  \\
 \cmidrule(lr) {2-5} \cmidrule(lr) {6-9} \cmidrule(lr) {10-13}
& \multicolumn{2}{c}{Estimate} & \multicolumn{2}{c}{$t$-ratio} & \multicolumn{2}{c}{Estimate} & \multicolumn{2}{c}{$t$-ratio} & \multicolumn{2}{c}{Estimate} & \multicolumn{2}{c}{$t$-ratio} \\
\midrule
Intercept  & -1&295 & -5&781*** & -0&428 & -1&690  
& -2&959 & -11&076***\\
$v1$ &  0&125 &  1&979*   & -0&087 & -1&210       
& -0&092 &  -1&131\\
$v2$ &  0&022 &  0&428    & -0&091 & -1&530    
&  0&106 &   1&581\\
$v3$ &  0&158 &  2&744*** & -0&108 & -1&687   
&  0&068 &   0&903\\
$v4$ & -0&274 & -4&641*** & 0&562 &   7&563***       
&  0&126 &   1&526\\
$v5$ & -0&011 & -0&063    & -0&260 & -1&300     
&  0&437 &   2&041*\\
$v6$ & -0&050 & -0&263    & -0&393 & -1&881         
&  0&103 &   0&455\\
$v7$ & -0&173 & -2&758*** & -0&038 & -0&533      
& -0&163 &  -2&028*\\
$v8$ &  0&192 &  2&097*   & -0&040 & -0&389      
& -0&072 &  -0&617\\
$v9$ &  0&578 &  7&123*** & -0&552 & -5&437***   
&  3&421 &  31&255***\\
$v10$&  0&773 &  8&812*** & -0&696 & -6&916***      
&  1&585 &  17&356***\\
$v11$& -0&000 & -0&005    &  0&072 &  0&784  
&  0&362 &   3&840***\\
\midrule
& \multicolumn{4}{c}{$W_1$} & \multicolumn{4}{c}{$W_2$} \\
 \cmidrule(lr) {2-5} \cmidrule(lr) {6-9}
& \multicolumn{2}{c}{Estimate} & \multicolumn{2}{c}{$t$-ratio} & \multicolumn{2}{c}{Estimate} & \multicolumn{2}{c}{$t$-ratio}
\\
Intercept & -1&243 & -40&006*** & -1&404 & -35&592*** \\
$v9$      & \multicolumn{4}{c}{}&  0&686 &  13&072*** \\
$\phi$    &  0&690 &  10&253*** &  0&824 &  12&159*** \\
\bottomrule
\end{tabular}\\
\small Signif. codes: 0 `***' 0.001 `**' 0.01 `*' 0.05.
\label{MZIH1}
\end{table}

\begin{table}[htbp]
\setlength{\belowcaptionskip}{10pt}
\caption{Estimation results under the Type I MZMH model.}
\centering
\begin{tabular}{lr@{.}lr@{.}lr@{.}lr@{.}lr@{.}lr@{.}l}
\toprule
& \multicolumn{4}{c}{$\pi'_0$} & \multicolumn{4}{c}{$\pi_1$} 
& \multicolumn{4}{c}{$\pi_2$}  \\
 \cmidrule(lr) {2-5} \cmidrule(lr) {6-9} \cmidrule(lr) {10-13}
& \multicolumn{2}{c}{Estimate} & \multicolumn{2}{c}{$t$-ratio} & \multicolumn{2}{c}{Estimate} & \multicolumn{2}{c}{$t$-ratio} & \multicolumn{2}{c}{Estimate} & \multicolumn{2}{c}{$t$-ratio} \\
\midrule
Intercept  & -2&473 & -27&212*** & -0&203 & -0&722  
& -2&724 & -9&367***\\
$v1$ &  0&047 &  1&562   & -0&095 & -1&120       
& -0&109 & -1&147\\
$v2$ &  0&026 &  1&111    & -0&109 & -1&612   
&  0&069 &  0&931\\
$v3$ &  0&134 &  5&132*** & -0&134 & -1&790   
&  0&046 &  0&536\\
$v4$ & -0&020 & -0&677    &  0&562 &  6&895***       
&  0&165 &  1&863\\
$v5$ &  0&036 &  0&472    & -0&312 & -1&391   
&  0&300 &  1&276\\
$v6$ & -0&144 & -1&816    & -0&408 & -1&734         
&  0&012 &  0&049\\
$v7$ & -0&205 & -6&884*** & -0&069 & -0&842      
& -0&182 & -1&972*\\
$v8$ &  0&127 &  2&934**  & -0&087 & -0&722      
& -0&112 & -0&831\\
$v9$ &  1&354 & 44&719*** & -0&579 & -5&518***     
&  3&386 & 30&065***\\
$v10$&  0&775 & 29&267*** & -0&731 & -7&056***      
&  1&547 & 16&541***\\
$v11$&  0&128 &  4&034*** & -0&063 & -0&615  
&  0&292 &  2&769**\\
\midrule
& \multicolumn{4}{c}{$W_1$} & \multicolumn{4}{c}{$W_2$} \\
 \cmidrule(lr) {2-5} \cmidrule(lr) {6-9}
& \multicolumn{2}{c}{Estimate} & \multicolumn{2}{c}{$t$-ratio} & \multicolumn{2}{c}{Estimate} & \multicolumn{2}{c}{$t$-ratio}
\\
Intercept & -1&243 & -40&006*** & -1&404 & -35&592*** \\
$v9$      & \multicolumn{4}{c}{}&  0&686 &  13&072*** \\
$\phi$    &  0&690 &  10&253*** &  0&824 &  12&159*** \\
\bottomrule
\end{tabular}\\
\small Signif. codes: 0 `***' 0.001 `**' 0.01 `*' 0.05.
\label{MZTH1}
\end{table}

\subsection{Ratemaking}
In this subsection, the analysis of several fitted models for ratemaking is conducted. We select five representative risk profiles: Best, Good, Average, Bad and Worst, and then calculate the means and variances of $Z_{1}+Z_{2}$ under each model for these selected risk profiles. The covariate information for the five different risk profiles is given in Table \ref{profile}. The formulas for these calculations can be found in the appendix.

\begin{table}[htbp]
\setlength{\belowcaptionskip}{10pt}
\centering
\caption{Covariate information for five different risk profiles.}
\begin{tabular}{cccccccccccc} \toprule
Profile & $v1$  & $v2$ & $v3$ & $v4$ & $v5$ & $v6$  & $v7$ & $v8$ & $v9$  & $v10$ & $v11$   \\
\midrule
Best    & 0 & 1 & 0 & 0 & 0 & 1 & 1 & 0 & 0 & 0 & 0\\
Good    & 0 & 0 & 1 & 0 & 0 & 1 & 0 & 0 & 0 & 0 & 1\\
Average & 0 & 1 & 0 & 0 & 0 & 1 & 1 & 0 & 0 & 1 & 1\\
Bad     & 0 & 0 & 0 & 1 & 0 & 1 & 1 & 0 & 1 & 0 & 0\\
Worst   & 1 & 1 & 1 & 0 & 1 & 0 & 0 & 1 & 1 & 0 & 1\\
\bottomrule
\end{tabular}\\
\label{profile}
\end{table}

Table \ref{ratemaking} shows the results for the five profiles under the ten fitted models. We observe that the predicted variances are always greater than the predicted means regardless of the profiles, indicating that overdispersion is detected in all models. The differences in the mean and variance estimates for the first three profiles are not noticeable between each model. However, the differences get more pronounced for the last two profiles. For the ``Bad'' profile, the five inflated models consistently lead to smaller means and variances than their modified counterparts. Both in the group of zero-inflated and zero-modified models, the two hurdle models give the largest variances. However, the opposite rule is shown for the ``Worst'' profile. Generally speaking, models derived from the negative binomial distribution produce more significant variance than those derived from the Poisson distribution due to the extra dispersion parameter in the negative binomial. The differences between the Type I MZIP and Type II MZIP, the Type I MZMP and Type II MZMP are only minor. The reason is that the dependence parameter $\lambda_0$ is very close to zero.

\begin{table}[htbp]
\setlength{\belowcaptionskip}{10pt}
\caption{Comparison of a priori ratemaking for different profiles under different models.}
\centering
\begin{tabular}{ccccccccccc}
\toprule
Model &\multicolumn{2}{c}{Best} & \multicolumn{2}{c}{Good} & \multicolumn{2}{c}{Average} & \multicolumn{2}{c}{Bad} & \multicolumn{2}{c}{Worst} \\
 \cmidrule(lr) {2-3} \cmidrule(lr) {4-5} \cmidrule(lr) {6-7} \cmidrule(lr) {8-9} \cmidrule(lr) {10-11}
  & Mean & Var & Mean & Var & Mean & Var & Mean & Var & Mean & Var\\
\midrule
Type I MZIP   & 0.078 & 0.120 & 0.110 & 0.162 & 0.186 & 0.289 & 0.286 & 0.538 & 0.655 & 1.096 \\
Type I MZINB  & 0.078 & 0.124 & 0.109 & 0.167 & 0.186 & 0.295 & 0.284 & 0.549 & 0.657 & 1.162 \\
Type I MZIH   & 0.077 & 0.126 & 0.113 & 0.185 & 0.184 & 0.300 & 0.303 & 0.647 & 0.618 & 1.060 \\
Type II MZIP  & 0.078 & 0.121 & 0.110 & 0.163 & 0.186 & 0.291 & 0.284 & 0.534 & 0.656 & 1.098 \\
Type II MZINB & 0.074 & 0.122 & 0.114 & 0.178 & 0.187 & 0.312 & 0.291 & 0.602 & 0.612 & 1.318 \\
\midrule
Type I MZMP   & 0.078 & 0.121 & 0.115 & 0.169 & 0.185 & 0.288 & 0.313 & 0.584 & 0.629 & 1.065 \\
Type I MZMNB  & 0.078 & 0.126 & 0.115 & 0.175 & 0.185 & 0.294 & 0.311 & 0.598 & 0.633 & 1.132 \\
Type I MZMH   & 0.077 & 0.126 & 0.119 & 0.194 & 0.184 & 0.300 & 0.336 & 0.721 & 0.589 & 1.014 \\
Type II MZMP  & 0.078 & 0.122 & 0.115 & 0.170 & 0.186 & 0.290 & 0.312 & 0.581 & 0.629 & 1.067 \\
Type II MZMNB & 0.078 & 0.127 & 0.115 & 0.180 & 0.185 & 0.308 & 0.312 & 0.665 & 0.634 & 1.311 \\
\bottomrule
\end{tabular}
\label{ratemaking}
\end{table}

\subsection{Multivariate zero-deflation}
To demonstrate the flexibility of our proposed multivariate zero-modified models to deal with the multivariate zero-deflation feature, we only reserve 5\% of common zeros randomly from our original data set. The processed data then contains 13,461 policyholders. The overall Pearson's correlation coefficient between these the two margins becomes negative(-0.144). Thus, the five multivariate zero-inflated models which assume positive correlations between each margins fail in this case. However, the proposed five multivariate zero-modified models can still handle this feature. The fitting results for the five modified models without covariates incorporated are listed in Table \ref{comparison3}. It is noted that $\hat{\pi}_0<\hat{\pi}'_0$ in all five models, suggesting a multivariate zero-deflation phenomenon. Again, the Type I MZMH model outperforms the other four candidates according to several information criteria. This confirms the benefit of separate treatment of zeros and positive part of the margin in a hurdle model.

\begin{table}[htbp]
\setlength{\belowcaptionskip}{10pt}
\caption{Comparison results of five multivariate zero-modified models without covariates incorporated.}
\centering
\begin{tabular}{ccccccc}
\toprule
Model & Parameters & LogLik & AIC & BIC & $\hat{\pi}_0$ & $\hat{\pi}'_0$ \\
\midrule
Type I MZMP   & 3 & -26,309.81 & 52,625.61 & 52,648.13 & 0.582 & 0.736 \\
Type I MZMNB  & 5 & -25,780.31 & 51,570.62 & 51,608.15 & 0.406 & 0.736 \\
Type I MZMH   & 7 & -25,767.25 & 51,548.49 & 51,601.05 & 0.416 & 0.736 \\
Type II MZMP  & 4 & -26,309.81 & 52,627.61 & 52,657.64 & 0.582 & 0.736 \\
Type II MZMNB & 4 & -25,989.73 & 51,987.47 & 52,017.50 & 0.202 & 0.736 \\
\bottomrule
\end{tabular}
\label{comparison3}
\end{table}

\section{Concluding remarks}

In this article, we discussed two types of models to deal with the multivariate zero-inflation feature, which is common to many automobile insurance data sets. The first type is the multivariate zero-inflated model, where we used one Bernoulli random variable to add common zeros to the original multivariate distribution. The second type is called the multivariate zero-modified model, where we used one Bernoulli random variable to truncate all common zeros and then used a second Bernoulli variable to add back the desired number of zeros. Models constructed from the second method can flexibly deal with both multivariate zero-inflation and zero-deflation phenomena. Several typical cases were presented with their corresponding inference procedures. 

These models were then compared comprehensively based on a data set from automobile insurance. The analysis showed that the two types of models were equivalent in modeling multivariate zero-inflated data when no covariates were incorporated. However, this equivalence no longer existed when covariates were introduced in different model components. With covariates in place, the comparison results showed consistently better performance of the multivariate zero-inflated models than their zero-modified counterparts for this particular data set. Overall, the Type I MZIH and Type I MZMH models outperformed their multivariate zero-inflated and multivariate zero-modified alternatives. It is evident that the enhanced performance is derived from the hurdle structure embedded in the two models. Next, the ten models were compared through {\em a priori} ratemaking analysis. As the risk of the profile went up, the differences between the mean and variance estimates of the total number of claims became more prominent. 

To demonstrate that the multivariate zero-modified model has wider applicability than the multivariate zero-inflated model, we conducted some further numerical analysis using a subset of the automobile insurance data set that shows multivariate zero-deflation features. The numerical results confirm that the multivariate zero-modified model can deal with such type of data properly.

\section*{Appendix}
\begin{appendix}
\section{Distributional properties}
\subsection{Multivariate zero-inflated models}

\begin{table}[htbp]
\setlength{\belowcaptionskip}{10pt}
\caption{The expectation and variance of $Z_j$, $j=1,\ldots,m$, under five multivariate zero-inflated models.}
\centering
\begin{tabular}{ccc}
\toprule
Model & $\mbox{E}(Z_j)$ & $\mbox{Var}(Z_j)$ \\
\midrule
Type I MZIP   & $\pi_0\lambda_{j}$ 
              & $\pi_0\lambda_j+\pi_0(1 -\pi_0)\lambda_j^2$ \\
\specialrule{0em}{3pt}{3pt}
Type I MZINB  & $\pi_0\lambda_{j}$ 
              & $\pi_0\lambda_j+\pi_0(1+1/\phi_j      -\pi_0)\lambda_j^2$ \\
\specialrule{0em}{3pt}{3pt}
Type I MZIHNB & $\pi_0 \pi_j (\lambda_j+1)$ 
              & $\pi_0\pi_j(\lambda_j+\lambda_j^2/\phi_j)+\pi_0\pi_j(1-\pi_0\pi_j)(\lambda_j +1)^2$ \\
\specialrule{0em}{3pt}{3pt}              
Type II MZIP  & $\pi_0(\lambda_{j}+\lambda_0)$ 
              & $\pi_0(\lambda_{j}+\lambda_0)+\pi_0(1 -\pi_0)(\lambda_{j}+\lambda_0)^2$ \\
\specialrule{0em}{3pt}{3pt}              
Type II MZINB & $\pi_0\lambda_{j}$ 
              & $\pi_0\lambda_j+\pi_0(1+1/\phi-\pi_0) \lambda_j^2$\\
\bottomrule
\end{tabular}
\end{table}

\begin{table}[htbp]
\setlength{\belowcaptionskip}{10pt}
\caption{The covariance between $Z_j$ and $Z_{j'}$, $j,j'=1,\ldots,m$, $j \neq j'$, under five multivariate zero-inflated models.}
\centering
\begin{tabular}{cc}
\toprule
Model & $\mbox{Cov}(Z_j,Z_{j'})$\\
\midrule
Type I MZIP   & $\pi_0(1-\pi_0)\lambda_j\lambda_{j'}$\\
\specialrule{0em}{3pt}{3pt}
Type I MZINB  & $\pi_0(1-\pi_0)\lambda_j\lambda_{j'}$ \\
\specialrule{0em}{3pt}{3pt}
Type I MZIHNB & $\pi_0(1-\pi_0)\pi_j \pi_{j'}                               (\lambda_j+1)(\lambda_{j'}+1)$\\
\specialrule{0em}{3pt}{3pt}
Type II MZIP  & $\pi_0\lambda_0+\pi_0(1-\pi_0)(\lambda_{j}+                 \lambda_0)(\lambda_{j'}+\lambda_0)$ \\
\specialrule{0em}{3pt}{3pt}
Type II MZINB & $\pi_0(1+1/\phi-\pi_0)\lambda_j\lambda_{j'}$ \\
\bottomrule
\end{tabular}
\end{table}

\subsection{Multivariate zero-modified models}
\begin{table}[htbp]
\setlength{\belowcaptionskip}{10pt}
\caption{The expectation of $Z_j$, $j=1,\ldots,m$, under five multivariate zero-modified models.}
\centering
\begin{tabular}{cccc}
\toprule
Model & $\pi_0$ & $\varphi$ & $\mbox{E}(Z_j)$ \\
\midrule
Type I MZMP   & $1-\exp(-\sum_{j=1}^m \lambda_j)$ 
              & $\pi'_0/\pi_0$ 
              & $\varphi\lambda_{j}$ \\
\specialrule{0em}{3pt}{3pt}
Type I MZMNB  & $1-\prod_{j=1}^m [\phi_j/(\lambda                            _j+\phi_j)]^{\phi_j}$ 
              & $\pi'_0/\pi_0$ 
              & $\varphi\lambda_{j}$ \\
\specialrule{0em}{3pt}{3pt}              
Type I MZMHNB & $1-\prod_{j=1}^m(1-\pi_j)$ 
              & $\pi'_0/\pi_0$ 
              & $\varphi \pi_j (\lambda_j+1)$ \\
\specialrule{0em}{3pt}{3pt}
Type II MZMP  & $1-\exp(-\sum_{j=0}^m \lambda_j)$ 
              & $\pi'_0/\pi_0$ 
              & $\varphi(\lambda_{j}+\lambda_0)$\\
\specialrule{0em}{3pt}{3pt}
Type II MZMNB & $1- [\phi/(\sum_{j=1}^m                                     \lambda_j+\phi)]^{\phi}$ 
              & $\pi'_0/\pi_0$ 
              & $\varphi\lambda_{j}$\\
\bottomrule
\end{tabular}
\label{expectation2}
\end{table}

\begin{table}[htbp]
\setlength{\belowcaptionskip}{10pt}
\caption{The variance of $Z_j$, $j=1,\ldots,m$, unfer five multivariate zero-modified models.}
\centering
\begin{tabular}{ccc}
\toprule
Model  & $\mbox{Var}(Z_j)$ \\
\midrule
Type I MZMP   & $\varphi\lambda_j+\varphi(1-\varphi)                        \lambda_j^2$ \\
\specialrule{0em}{3pt}{3pt}              
Type I MZMNB  & $\varphi\lambda_j+\varphi(1+1/\phi_j-                       \varphi)\lambda_j^2$ \\
              
\specialrule{0em}{3pt}{3pt}              
Type I MZMHNB & $\varphi\pi_j(\lambda_j+\lambda_j^2/\phi_j)+                 \varphi\pi_j(1-\varphi\pi_j)(\lambda_j                      +1)^2$ \\
\specialrule{0em}{3pt}{3pt}              
Type II MZMP  & $\varphi(\lambda_{j}+\lambda_0)+\varphi(1-                  \varphi)(\lambda_{j}+\lambda_0)^2$ \\
\specialrule{0em}{3pt}{3pt}               
Type II MZMNB & $\varphi\lambda_j+\varphi(1+1/\phi-\varphi)                 \lambda_j^2$\\
\bottomrule
\footnotesize{$\varphi$ can be referred to Table \ref{expectation2}}
\end{tabular}
\end{table}

\begin{table}[htbp]
\setlength{\belowcaptionskip}{10pt}
\caption{The covariance between $Z_j$ and $Z_{j'}$, $j,j'=1,\ldots,m$, $j \neq j'$, under five multivariate zero-modified models.}
\centering
\begin{tabular}{ccc}
\toprule
Model  & $\mbox{Cov}(Z_j,Z_{j'})$ \\
\midrule
Type I MZMP   & $\varphi(1-\varphi)\lambda_j\lambda_{j'}$\\
\specialrule{0em}{3pt}{3pt}
Type I MZMNB  & $\varphi(1-\varphi)\lambda_j\lambda_{j'}$ \\
\specialrule{0em}{3pt}{3pt}
Type I MZMHNB & $\varphi(1-\varphi)\pi_j \pi_{j'}                               (\lambda_j+1)(\lambda_{j'}+1)$\\
\specialrule{0em}{3pt}{3pt}
Type II MZMP  & $\varphi\lambda_0+\varphi(1-\varphi)(\lambda_{j}+                 \lambda_0)(\lambda_{j'}+\lambda_0)$ \\
\specialrule{0em}{3pt}{3pt}
Type II MZMNB & $\varphi(1+1/\phi-\varphi)\lambda_j\lambda_{j'}$ \\
\bottomrule
\footnotesize{$\varphi$ can be referred to Table \ref{expectation2}}
\end{tabular}
\end{table}

\end{appendix}

\end{document}